\documentclass[twocolumn,trackchanges]{aastex631}

\usepackage{graphicx}	
\usepackage{amsmath}	
\usepackage{blindtext}
\usepackage{subcaption}
\usepackage{calc}

\usepackage{hyperref}
\hypersetup{
    colorlinks=true,
    linkcolor=black,
    filecolor=blue,      
    urlcolor=blue,
    pdfpagemode=FullScreen,
    }

\begin{document}

\title{\Large Dark matter distribution in Milky Way-analog galaxies}

\author[0009-0005-1424-6604, $^\dagger$]{Natanael G. de Isídio}\email{$\dagger$ isidio@astro.ufrj.br}
\affiliation{Valongo Observatory, Federal University of Rio de Janeiro, Ladeira de Pedro Antônio 43, Rio de Janeiro, RJ 20080090, Brazil}
\affiliation{National Radio Astronomy Observatory, Array Operations Center, Socorro, NM 87801, USA}

\author[0000-0003-3153-5123]{K. Menéndez-Delmestre}
\affiliation{Valongo Observatory, Federal University of Rio de Janeiro, Ladeira de Pedro Antônio 43, Rio de Janeiro, RJ 20080090, Brazil}

\author[0000-0003-2374-366X]{T. S. Gonçalves}
\affiliation{Valongo Observatory, Federal University of Rio de Janeiro, Ladeira de Pedro Antônio 43, Rio de Janeiro, RJ 20080090, Brazil}

\author[0000-0003-4675-3246]{M. Grossi}
\affiliation{Valongo Observatory, Federal University of Rio de Janeiro, Ladeira de Pedro Antônio 43, Rio de Janeiro, RJ 20080090, Brazil}

\author[0000-0003-1683-5443]{D. C. Rodrigues}
\affiliation{Physics departament, Federal University of Espírito Santo, Vitória, ES, Brazil}
\affiliation{Institute for Theoretical Physics, University of Heidelberg, Philosophenweg 16, D-69120, Heidelberg, Germany}

\author[0000-0001-7107-1744]{N. Garavito-Camargo}
\affiliation{Center for Computational Astrophysics, Flatiron Institute, 162 5th Ave, New York, NY 10010, USA}

\author[0000-0001-5539-0008]{A. Araujo-Carvalho}
\affiliation{Valongo Observatory, Federal University of Rio de Janeiro, Ladeira de Pedro Antônio 43, Rio de Janeiro, RJ 20080090, Brazil}

\author[0000-0001-5071-1343]{P. P. B. Beaklini}
\affiliation{National Radio Astronomy Observatory, Array Operations Center, Socorro, NM 87801, USA}

\author[0009-0007-3523-5140]{Y. Cavalcante-Coelho}
\affiliation{Valongo Observatory, Federal University of Rio de Janeiro, Ladeira de Pedro Antônio 43, Rio de Janeiro, RJ 20080090, Brazil}

\author[0000-0002-0620-136X]{A. Cortesi}
\affiliation{Valongo Observatory, Federal University of Rio de Janeiro, Ladeira de Pedro Antônio 43, Rio de Janeiro, RJ 20080090, Brazil}
\affiliation{Institute of Physics, Federal University of Rio de Janeiro, Av. Athos da Silveira Ramos 149, Rio de Janeiro, RJ 21941909, Brazil}

\author[0000-0002-9390-955X]{L. H. Quiroga-Nu\~nez}
\affiliation{Department of Aerospace, Physics, and Space Sciences, Florida Institute of Technology, 150 University Blvd, Melbourne, FL 32901, USA}

\author[0000-0002-8791-2138]{T. Randriamampandry}
\affiliation{Valongo Observatory, Federal University of Rio de Janeiro, Ladeira de Pedro Antônio 43, Rio de Janeiro, RJ 20080090, Brazil}
\affiliation{Department of Physics \& Astronomy, University of the Western Cape, Robert Sobukwe Rd, Bellville, 7535, South Africa}

\begin{abstract}
Our current understanding of how dark matter (DM) is distributed within the Milky Way halo, particularly in the solar neighborhood, is based on either careful studies of the local stellar orbits, model assumptions on the global shape of the MW halo, or from direct acceleration measurements.
In this work, we undertake a study of external galaxies, with the intent of providing insight to the DM distribution in MW-analog galaxies. 
For this, we carefully select a sample of galaxies similar to the MW, based on maximum atomic hydrogen (HI) rotational velocity ($v_{\rm max,HI}=200-280$~km~s$^{-1}$) and morphological type (Sab--Sbc) criteria.
With a need for deep, highly-resolved HI, our resulting sample is composed of 5 galaxies from the VIVA and THINGS surveys. 
To perform our baryonic analysis, we use deep Spitzer mid-IR images at 3.6 and 4.5~$\mu$m from the S$^4$G survey. 
Based on the dynamical three-dimensional modeling software $^{\rm{3D}}$\textsc{Barolo}, we construct rotation curves (RCs) and derive the gas and stellar contributions from the galaxy’s gaseous- and stellar-disks mass surface density profiles. 
Through a careful decomposition of their RCs into their baryonic (stars, gas) and DM components, we isolate the DM contribution by using an MCMC-based approach.
Based on the Sun’s location and the MW's R$_{25}$, we define the corresponding location of the solar neighborhood in these systems. We put forward a window for the DM density ($\rho_{dm}$~=~0.21--0.55~GeV~cm$^{-3}$) at these galactocentric distances in our MW analog sample, consistent with the values found for the MW's local DM density, based on more traditional approaches found in the literature.
\end{abstract}


%
%
%
%
\section{Introduction} \label{sec:introduction}
It has been more than 50 years since \cite{1970ApJ...159..379R} noticed that the rotation curve (RC)\footnote{Throughout this paper, the term ‘rotation curve’ is used to refer to the measured rotation velocity of HI. For the velocity corresponding the mass distribution of a particular component of a galaxy we use the term ‘circular velocity’, since $v_{\rm c}^2(r)=r~\frac{\partial \Phi}{\partial r}$.} of M31 exhibits unexpected rotation dynamics according to Newtonian physics. Despite decades of research and relevant progress in numerical simulations \citep[e.g.,][]{2008Natur.456...73S,2009MNRAS.398.1150B,2010MNRAS.402...21N}, dark matter (DM) remains one of the most important open challenges in contemporary astronomy.
In the standard $\Lambda$-cold dark matter ($\Lambda$CDM) cosmological model, DM is the primary component of the mass content of the universe, with recent estimates indicating that DM comprises $\sim$85\% of all the mass \citep[e.g.,][]{2016A&A...594A..13P,2020ApJ...901...90A}.
Despite its predominance, the nature of DM is still poorly understood and many questions regarding its properties and behavior remain unanswered.

Uncovering the nature of the DM particle through the direct detection of DM interactions is the primary objective of many high-energy research facilities, including the LUX-ZEPLIN Experiment \citep[LZ;][]{2015arXiv150902910T,2020PhRvD.101e2002A,2022arXiv220703764A}, and the XENON Experiment \citep[][]{2017PhRvL.119r1301A}. 
Both LZ and XENON aim to directly detect DM in the form of weakly interacting massive particles \citep[WIMPs;][]{1985NuPhB.253..375S}.
Readers may refer to \citet[][]{2018EPJC...78..203A} for a recent review on WIMPs.
This kind of experiments have been striving to push the boundaries in the quest to detect the DM particle.
One important factor for the success of WIMP-based experiments is to have an accurate estimate of the local DM density (at the Sun's position, at $\sim$8 kpc from the Galactic centre). This reference value allows generating predictions of the expected number of detectable particles in direct-detection experiments \citep[][]{1996APh.....6...87L}.
The estimate for the local DM density in the solar neighborhood was first estimated over a century ago by \citet[][]{1922ApJ....55..302K}, who used the vertical motions of stars near the Sun.
However, with the gradual increase in both the quantity and quality of observational data, local estimates have been refined \citep[e.g.,][]{2014JPhG...41f3101R}.
There are two main kinematic approaches to determine the local DM density: measuring vertical motion of stars in the solar neighbourhood \citep[e.g.,][]{1922ApJ....55..302K,1922MNRAS..82..122J,1932BAN.....6..249O,1987A&A...180...94B,1998A&A...329..920C,2004MNRAS.352..440H,2012ApJ...756...89B,2013ApJ...772..108Z}, and using global assumptions about the shape of the MW DM halo \citep[e.g.,][]{1992AJ....103.1552M,1998MNRAS.294..429D,2009PASJ...61..153S,2010A&A...509A..25W,2010JCAP...08..004C,2011MNRAS.414.2446M}.
In the first approach, estimates from vertical motions rely on local conditions of stellar kinematics that carry significant uncertainties \citep[e.g.,][]{2011JCAP...11..029I,2014JPhG...41f3101R,2020Galax...8...37S}. 
In the second approach, despite the advantage of global measures presenting small error bars, estimates are strongly biased due to assumptions regarding the distribution of DM in the MW \citep[e.g.,][]{2011JCAP...11..029I,2014JPhG...41f3101R}.
There have been recent attempts to bring these two approaches together by modelling the phase-space distribution of stars in larger volumes around the solar neighborhood \citep[][]{2013ApJ...779..115B}.
Also, recent works using extreme-precision time-series observations that provide direct acceleration measurements have been gaining increasing visibility \citep[e.g.,][]{2019PASA...36...38S,2020ApJ...902L..28C,Chakrabarti+21,2022ApJ...928L..17C}. For instance, \citet[][]{Chakrabarti+21} have used high-precision pulsar timing measurements to measure Galactic acceleration and provide their own local DM density estimate.
We note, however, that the structure, extension and mass of the MW are still challenging to observationally quantify robustly due to the inherent challenge of globally studying the system in which we are located.
We here set out to use external galaxies in an attempt to go around some of these challenges.

Even when the intention is to study the local DM density, the choice of mass distribution models remains a point that must be carefully addressed. A large number of studies \citep[e.g.,][]{2016ApJ...817...84K,2017MNRAS.464...65P,2019A&A...626A..56P,2020Galax...8...37S,2022MNRAS.514.3329M}, relying on different techniques, converge onto similar results: the baryonic contribution decreases sharply with galactocentric distance, while the contribution of DM tends to become increasingly dominant in the outskirts of galaxies.
Despite this general consensus, an important open question concerns the galactocentric distance at which DM becomes dominant over the baryonic component \citep[e.g.,][]{2021A&A...649A.119P,2021ApJ...922..143P,2023arXiv230906390B}.
Recent work by \citet[][]{2021A&A...649A.119P} suggests that the DM content becomes comparable to that of baryons around $\sim$2--3 R$_e$.
The way in which DM is distributed within galaxies has been a topic of great interest, as it relates to the formation history of galaxies over cosmic time \citep[e.g.,][]{2019A&ARv..27....2S,2020Univ....6..107D,2021PDU....3200826B,2022ApJ...934...43S,2023MNRAS.tmp.1363Y}.

The RCs of disk galaxies are a crucial tool to probe for the DM content and DM radial distribution \citep[e.g.,][]{2001ARA&A..39..137S,2008AJ....136.2563W,2008AJ....136.2648D,2009AJ....138.1741C,2017PASJ...69R...1S,2022MNRAS.514.3329M,2022vhow.confE..18D,2023MNRAS.518.6340D}.
These curves provide precise information about motions and internal kinematics of gas within these systems. By decomposing a galaxy's RC into its different components (e.g., stellar disk, gaseous disk, and DM halo), we can distinguish how much each component (stars, gas, DM) is contributing to the total observed mass. RCs also provide a means to infer and compare the evolutionary trajectory of the system, describing the role that interactions have played in its structure over cosmic time \citep[e.g.,][]{2001ARA&A..39..137S,2019A&A...622A.197A,2019PhRvD.100h3518S}.
RCs based on the orbital velocity of the atomic hydrogen gas (hereafter HI) are particularly important to understand the distribution of DM, as the observed extension of the HI distribution is often $\sim$2-3 times larger than the spatial extent of the stellar light, as seen in optical and infrared bands \citep[e.g.,][]{2008AJ....136.2563W,2009AJ....138.1741C}.


In this work, we combine HI spectroscopy with mid-infrared (mid-IR) imaging at 3.6 and 4.5 $\mu$m to investigate the DM distribution within MW-analog systems.
The mid-IR insight in these bands allows us a detailed view of the stellar mass distribution as it is the older, low-mass stars, responsible for the bulk of the stellar mass in the galaxy that dominates at these wavebands \citep[e.g.,][]{2012AJ....143..139E,2015ApJS..219....5Q}.
Based on the decomposition of the HI-derived RCs for these systems, we isolate the DM component and provide estimates for the density of DM in the region corresponding to the location of the solar neighborhood in these systems.
By drawing a parallel with our own MW, this distinct approach based on external galaxies provides a potentially useful insight to the range of valid values for the local DM density, $\rho_{dm}$.


This paper is structured as follows. In Section \ref{sec:sample}, we present the sample selection criteria of our MW-analog galaxies and the data (mid-infrared imaging and HI spectroscopy) used throughout this work. 
Section \ref{sec:methods} goes into the details of the methods used to explore the DM distribution in this work.
In Section \ref{sec:results}, we present and discuss our main results. 
Finally, in Section \ref{sec:discussion} we present a brief summary of the work. 

Throughout this paper, we employ a flat cosmology based on the standard $\Lambda$CDM model, with cosmological constraints set to $H_0$~=~(68.7~$\pm$~3.1) km~s$^{-1}$~Mpc$^{-1}$, $\Omega_m=0.3$, $\Omega_{\Lambda}=0.7$, and a cosmological baryon fraction of $f_b = \Omega_b/\Omega_m = 0.187$ \citep[][]{2020A&A...641A..10P}.

%
%
%
%
\section{Sample and data}
\label{sec:sample}
We select a sample of galaxies whose physical characteristics are similar to those of the MW.
Our galaxies were selected based on their HI velocity, morphology and inclination, respecting the cut-off criteria described in the following subsection.
We build RCs for each galaxy in our sample based on publicly available HI data and, with the aim of analysing separately the DM and baryonic components, we further exploit available mid-IR datasets.

\subsection{Sample selection criteria}\label{subsec:sample criteria}
Our main interest is to analyze DM halos similar to that of the MW.
In order to identify these halos, we consider the HI velocity as a cut-off criterion.
Recent studies show that the maximum HI velocity in the MW converges to values within the range of 220-240~km~s$^{-1}$ \citep[e.g.,][]{2016ApJ...832..159R,2020Galax...8...37S}. In this work we select galaxies with a maximum HI velocity within the range of 200-280~km~s$^{-1}$.

In an interest to probe not only galaxies with a DM halo similar to the MW, but also with a similar baryonic distribution, we further constrain our selection criteria to include galaxies with a similar morphological type to the MW.
Recent works have established that the MW is an SBbc barred spiral with typical Hubble T-type in the range of 2.0--3.0 \citep[][]{2014A&A...569A.125H,2017A&A...604A..72Q}. 
In this work we select galaxies with Hubble T-types within the range of 2.0--4.0, including galaxies with morphologies consistent with SAb/SABb/SBb, SAbc/SABbc/SBbc, and SAc/SABc/SBc galaxies.

Inclination is an important criterion for our sample selection. On the one hand, since our work is based on the analysis of HI rotation curves to trace the mass distribution in galaxies, we exclude galaxies with low inclination, where measured gas motions are dominated by the dispersion component. Focusing on galaxies with sufficient inclination allows us to probe the gas rotation in these systems.
On the other hand, since we also seek to select galaxies that share a similar morphology to that of the MW, we exclude galaxies with high inclinations that challenge a proper characterization of its morphology. 
Our inclination cut-off criterion excludes all galaxies that do not have an inclination between $35^{\circ} < i < 75^{\circ}$.

Finally, as our analysis relies on the availability of deep HI mapping to enable RC construction (see $\S$\ref{subsec:HI}), our final sample is composed of 5 MW-analog galaxies with publicly available HI data.
All galaxies in our sample have stellar masses within the range of $10^{10.6}-10^{11.1}$~$\textrm{M}_\odot$. Throughout this work, we adopt a stellar mass reference of $10^{10.78}$~$\textrm{M}_\odot$ for the MW \citep{2015ApJ...806...96L}.
The final sample can be found in Table \ref{tab:sample}.

\begin{table*}
\caption{Properties of the sample of Milky Way-analog galaxies}
\begin{center}
\begin{tabular}{ cccccccccccc }
\hline
    Galaxy & $\alpha_{2000}$ & $\delta_{2000}$ & Type & T-type & $\textrm{M}_\star$ & $\textrm{M}_{HI}$ & $v_{HI}$ & i & Distance & HI data \\
    & [hh:mm:ss] & [dd:mm:ss] & & & \multicolumn{2}{c}{[$\log_{10}(\textrm{M}/\textrm{M}_\odot)$]} &[km~s$^{-1}$] & [$^\circ$] & [Mpc] & \\
    (1) & (2) & (3) & (4) & (5) & (6) & (7) & (8) & (9) & (10) & (11)\\
\hline
\hline
    NGC 2903 & 09 32 10.1 & +21 30 03.0 & SBbc & 4.0 & $10.67$ & 9.66 & $207.7 \pm 4.5$  & 65.0 & $9.1 \pm 1.6$ & THINGS \\
    NGC 3521 & 11 05 48.6 & -00 02 09.1 & SABb & 4.0 & $11.03$ & 10.01 & $229.7 \pm 8.6$  & 73.0 & $12.1 \pm 2.8$ & THINGS \\
    NGC 4579 & 12 37 43.5 & +11 49 05.5 & SBb & 3.0 & $11.10$ & 8.80 & $277.1 \pm 4.4$ & 38.0 & $17.0 \pm 0.1$ & VIVA \\
    NGC 4698 & 12 48 22.9 & +08 29 14.6 & SAab  & 2.0 & $10.85$ & 9.27 & $255.2 \pm 27.4$ & 53.0 & $17.0 \pm 6.7$ &  VIVA\\
    NGC 5055 & 13 15 49.3 & +42 01 45.4 & SAbc & 4.0 & $10.78$ & 9.79 & $216.0 \pm 10.8$  & 59.0 & $8.3 \pm 1.9$ & THINGS \\
\hline
\end{tabular}
\end{center}
\label{tab:sample}
\bfseries{Notes:} \normalfont (1)~Galaxy identifier;
(2)~Right ascension and (3)~declination coordinates in J2000 taken from the NASA Extragalactic Database;
(4)~Morphological type taken from Hyperleda \citep[][]{2014A&A...570A..13M}; (5)~Hubble T-type morphology from \citet{1994AJ....108.2128C};
(6)~Stellar mass, taken from the S$^4$G value-added catalog \citep{2010PASP..122.1397S};
(7)~HI mass calculated from the integrated HI-line flux density using equation (\ref{eq:flux_HI_massa});
(8)~Maximum HI velocity derived from the rotation curve; 
(9)~Inclinations from \citet{2008AJ....136.2563W} and \citet{2009AJ....138.1741C} for THINGS and VIVA galaxies, respectively; 
(10)~Galaxy distance taken from \citet{2018ApJS..234...18B}. For VIVA galaxies we assume that all Virgo cluster galaxies are at a mean distance of 17 Mpc \citep[e.g.,][]{2012A&A...545A..75P,2018ApJS..234...18B};
(11)~HI survey where the data was sourced from.
\end{table*}

\subsection{HI data}
\label{subsec:HI}
Our HI data comes from two surveys: \textit{The VLA Imaging of Virgo in Atomic Gas} \citep[VIVA;][]{2009AJ....138.1741C} and \textit{The HI Nearby Galaxy Survey} \citep[THINGS;][]{2008AJ....136.2563W}. 
All the objects in the VIVA sample (53 late-type galaxies) were imaged using the Very Large Array (VLA) with typical spatial and spectral resolutions of $\sim$15$\arcsec$ and 10~km~s$^{-1}$, respectively. VIVA observations have typical column density sensitivities of $3-5~\times 10^{19}$~cm$^{-2}$.
As with VIVA, THINGS also used interferometry with VLA to obtain spatial resolutions of $\sim$~6$\arcsec$, resulting in linear resolutions of $\sim$100-500~pc for its entire sample of 34 nearby galaxies.
THINGS observations have typical column density sensitivities of $\sim 4~\times~10^{19}$~cm$^{-2}$ at 30$\arcsec$ resolution -- adequate for mapping the outer regions of galaxies \citep{2008AJ....136.2563W}. Readers may refer to \cite{2008AJ....136.2563W} and \cite[][]{2009AJ....138.1741C} for details on the surveys.

In this work, we study 2 MW-analog galaxies from VIVA and 3 MW analogs from THINGS.
Both VIVA and THINGS surveys provide HI flux, velocity and velocity dispersion maps in their databases.
For 45 galaxies in the VIVA data set, the reduced HI spectral line cubes are available\footnote{VIVA reduced HI spectral line cubes are available at \url{http://www.astro.yale.edu/cgi-bin/viva/observations.cgi}.}. For the THINGS sample, all the galaxies have reduced HI spectral line cubes available\footnote{THINGS reduced HI spectral line cubes are available at \url{https://www2.mpia-hd.mpg.de/THINGS/Data.html}.}.
We use the publicly-available cubes to derive HI gas masses for our sample and to construct an RCs for each galaxy in our sample (see $\S$\ref{sec:methods}). 

To derive HI gas masses for our sample, in M$_\odot$ units, we use the relation introduced by \citet{2010MNRAS.403..683C}:
\begin{equation}\label{eq:flux_HI_massa}
    \textrm{M}_{\rm HI}=\frac{2.356 \times 10^5}{1+z}\,\left[\frac{d_L(z)}{\textrm{Mpc}}\right]^2 \, \left(\frac{S_{\rm HI}}{\textrm{Jy~km~s}^{-1}}\right)
\end{equation}
where $d_L(z)$ is the luminosity distance to the galaxy at redshift $z$
, as derived from the HI spectrum, and $S_{\rm HI}$ is the integrated flux density of the HI line. This integrated flux density is calculated as follows:
\begin{equation}
    S_{HI} = \sum F_{\rm HI} \cdot \Delta v
\end{equation}
where $F_{\rm HI}$ is the HI flux and $\Delta v$ is the channel width (5.2 and 10.4~km~s$^{-1}$ for THINGS and VIVA data, respectively).
The total HI masses calculated are also shown in Table \ref{tab:sample}.
We note that our HI mass estimates are in agreement with the values presented by \citet[][]{2008AJ....136.2563W} and \citet[][]{2009AJ....138.1741C}.

\subsection{Mid-IR data} \label{subsec:S4G}

An integral part of this work is to characterize in detail the baryonic matter distribution of the sample, in order to properly isolate the DM component. 
Considering that the stellar mass represents $\sim$90\% of the total baryonic mass in nearby galaxies \citep{2010gfe..book.....M,2016AJ....151...94F}, we take a careful approach to characterize the stellar mass distribution in our galaxies. 
Using mid-IR fluxes has been shown to be a reliable and accurate method for estimating stellar masses in nearby galaxies \citep[e.g.,][]{2012AJ....143..139E,2023arXiv230105952J}. 
Motivated by this, we rely on the mid-IR emission traced by deep images from the \textit{Spitzer Survey of Stellar Structures in Galaxies} \citep[hereafter S$^4$G;][]{2010PASP..122.1397S,2013ApJ...771...59M,2015ApJS..219....5Q,https://doi.org/10.26131/irsa425,2022A&A...660A..69W}.
Consisting of 3.6 and 4.5~$\mu$m imaging with Spitzer’s Infrared Array Camera \citep[IRAC;][]{2004ApJS..154...10F} of 2353 galaxies down to surface mass densities of 1~M$_\odot~$pc$^{-2}$, S$^4$G is the deepest, largest and most-homogeneous mid-IR dataset for the nearby Universe.
The processing of all S$^4$G images was performed through a pipeline composed of several steps (P1-P4), as detailed in \citet{2010PASP..122.1397S}.
The resulting S$^4$G images have a pixel size of 0.75$\arcsec$ and a point spread function (PSF) of 1.66$\arcsec$ and 1.72$\arcsec$ for the 3.6 and 4.5~$\mu$m bands, respectively. 
We refer the reader to \citet[][]{2010PASP..122.1397S} and \citet[][]{2015ApJS..219....3M} for a full presentation of the survey and the details regarding the processing of the public dataset.

Although other works have opted to use S$^4$G’s  3.6~$\mu$m surface photometry and a fixed mass-to-luminosity (M/L) ratio to build mass models for nearby galaxies \citep[e.g.,][]{2016AJ....152..157L}, we adopt a different approach based on a careful consideration of the work by \citet{2015ApJS..219....5Q}. 
The authors point to the contamination that dust emission can bring into the 3.6~$\mu$m emission (to a level of $\sim$10-30\%) and provide a correction for this, by using the 4.5~$\mu$m band.
We compare the mass values found through the relation suggested by \citet{2015ApJS..219....5Q} using only mid-IR fluxes with those found using SED fitting across UV to radio wavelengths available in DustPedia catalog \citep[e.g.,][]{DustPedia,SED_Fitting,2022MNRAS.512.2728C}.
In particular, we note that when examining the region occupied by potential MW analogs, the method used to calculate stellar masses, whether considering only mid-IR fluxes or through SED fitting, yields similar results ($<$0.2~dex).
We derive the stellar mass of our galaxies using the equation introduced by \citet{2015ApJS..219....5Q}, which is used to convert galaxy flux densities at 3.6 and 4.5~$\mu$m into the galaxy’s stellar mass as follows:
\begin{equation}\label{eq:stellar_mass}
    \textrm{M}_\star = 10^{8.35}~F_{3.6\mu m}^{1.85}~F_{4.5\mu m}^{-0.85}~D^{2}
\end{equation}
where $\textrm{M}_\star$, $F$, and $D$ are given in M$_\odot$, Jy and Mpc units, respectively. We adopt an uncertainty of 0.2~dex on our stellar mass estimates as suggested in \citet{2015ApJS..219....5Q}.

To mitigate contamination from foreground stars within the object's field, we employ masks provided from the 3.6 and 4.5~$\mu$m images by the S$^4$G team \citep[see IRSA database\footnote{\url{http://irsa.ipac.caltech.edu/data/SPITZER/S4G}};][]{2010PASP..122.1397S}.

We observe that the relative discrepancies between the stellar and gas masses we show in Table~\ref{tab:sample} and the values available in the THINGS, VIVA, and S$^4$G databases \citep[][respectively]{2008AJ....136.2563W,2009AJ....138.1741C,2010PASP..122.1397S} do not exceed 30\%, and have no significant impact on the results we present here.

%
%
%
%
\section{Methods on HI rotation curves}
\label{sec:methods}
Our methodology is based on three main steps:  the three-dimensional kinematic modelling of the HI data cubes, the construction of rotation curves for our sample based on this modelling and their subsequent decomposition into the stellar, gas, and DM components. We complement this with an MCMC-based analysis of the DM component, where we estimate the DM contribution for the observed RCs and estimate the halo mass and concentration of our galaxy sample.
The entire analysis detailed in this section is applied to all galaxies within our sample.

\subsection{3D tilted-ring kinematical modelling}\label{subsec:barolo}
We construct the RC for each galaxy in our sample based on the publicly-available data cubes from VIVA and THINGS (see $\S\ref{subsec:HI}$), using the 3D-Based Analysis of Rotating Objects via Line Observations \citep[hereafter $^{\rm{3D}}$\textsc{Barolo};][]{2015MNRAS.451.3021D}.
The $^{\rm{3D}}$\textsc{Barolo} algorithm uses a tilted-ring model to provide precise analysis of spectroscopic observations of galaxies, allowing for a detailed mapping of the gas kinematics and geometry.
By using $^{\rm{3D}}$\textsc{Barolo}, we are able to derive the RCs for each object within our sample as well as the corresponding gas surface density, which allows us to decompose the RC into its different mass components (see $\S$\ref{subsec:mass_models}).
For this work we use the latest stable version\footnote{The software can be downloaded at \url{github.com/editeodoro/Bbarolo}} (1.6) of $^{\rm{3D}}$\textsc{Barolo}. 
Readers may refer to \citet[][]{2015MNRAS.451.3021D} for more details on the software.

By convolving the model velocity field with the beam size and shape, $^{\rm{3D}}$\textsc{Barolo} is able to accurately reproduce the observed velocity field and recover the true kinematics of the galaxy. The code also provides a number of diagnostic tools to evaluate the goodness-of-fit of the model and identify potential areas of improvement.
It has been successfully applied to a number of observational studies, providing valuable insights into the kinematics and dynamics of galaxies across a range of morphologies in both nearby \citep[e.g.,][]{2022MNRAS.514.3329M,2022ApJ...934..173S,2023MNRAS.518.6340D} and high-redshift galaxies \citep[e.g.,][]{2021MNRAS.503.1753S,2021MNRAS.507.3540J,2023MNRAS.521.1045R}, where the analysis of their kinematic properties provides crucial information on their evolution and formation processes. 

In this work we feed $^{\rm{\rm{3D}}}$\textsc{Barolo} with various input parameters -- including distance and inclination, while keeping many others as free parameters for a more accurate determination. 
These free parameters include the rotational velocity (VROT), dispersion velocity (VDISP), and position angle (PA). 
The distances for all 5 galaxies are based on \citet{2018ApJS..234...18B}, where all galaxy members of the Virgo cluster are set to a distance of 17~Mpc. 
The inclinations were taken from \citet[][]{2008AJ....136.2563W} and \citet[][]{2009AJ....138.1741C}, for the VIVA and THINGS galaxies, respectively.

\begin{figure}
\centering
\includegraphics[width=0.47\textwidth]{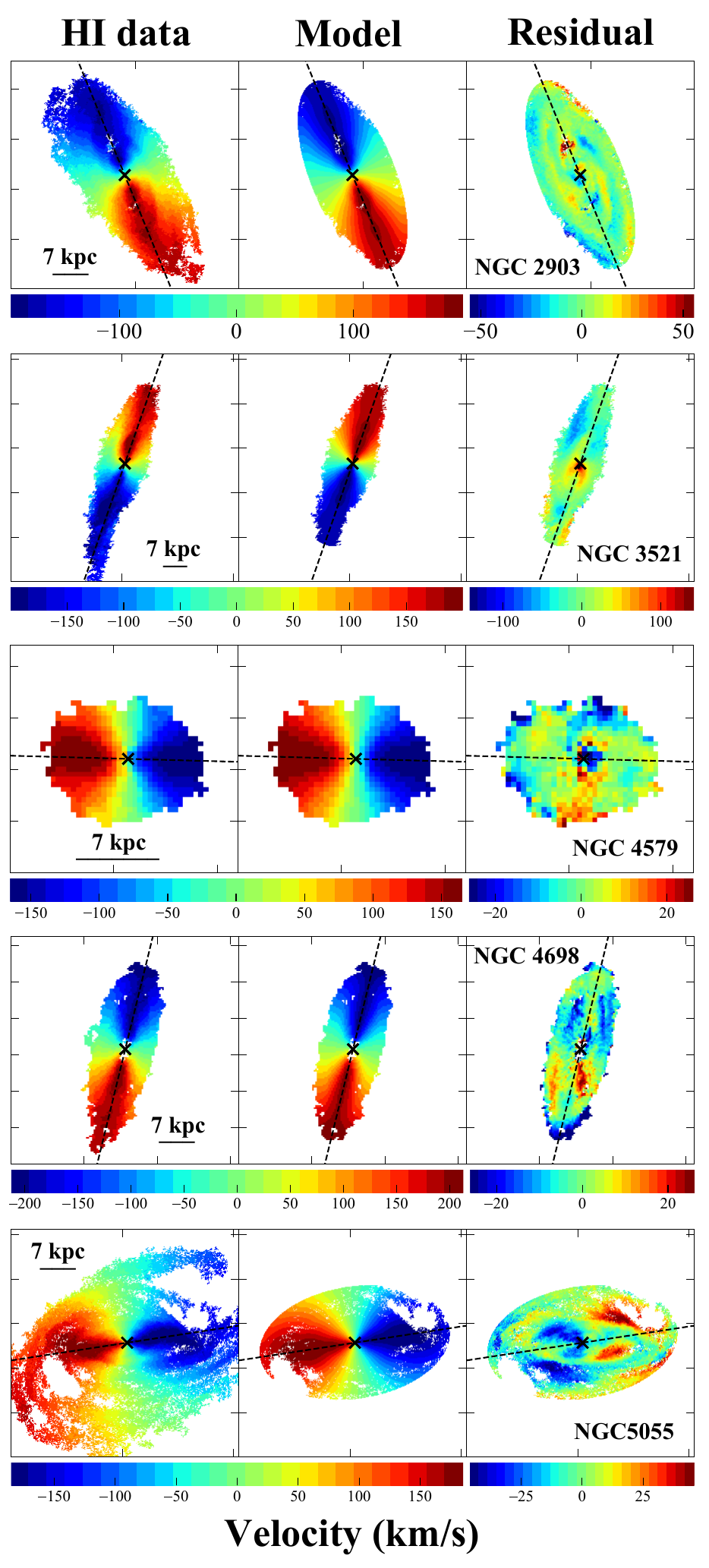}
\caption{Velocity maps of the HI data (left column) and model (middle column) with the corresponding residuals (right column) for our MW-analog sample. Each row represents the data for each of our 5 MW analogs. Our model is derived based on the $^{\rm{\rm{3D}}}$\textsc{Barolo} analysis (see $\S$\ref{subsec:barolo}). Black dashed line shows the major axis of the galaxy, with the ‘X' symbol marking the galaxy’s kinematic centre. Velocities are shown in the galaxy’s rest frame. For all galaxies in this work, high-resolution images are available in the \href{https://doi.org/10.5281/zenodo.11396279}{Zenodo repository}.}
    \label{fig:plot_Barolo}
\end{figure}

We use a masking mode (via ‘SEARCH’ method) to eliminate contaminants within the cube and identify genuine line emitting regions.
We minimize the discrepancy between the modeled and observed data using the chi-squared metric to achieve the best-fit solution.
We use the WFUNC parameter to weight the fit based on the absolute cosine of the azimuthal angle (i.e., $|\cos(\theta)|$), in order to find the best-fit solution.
To define the convergence criterion for the fitting algorithm, we applied the tolerance default value of 0.003 ensuring that the modeled and observed data reach a discrepancy below the specified threshold suggested by \citet{2015PhDT.......216D}.
We conducted the analysis on both sides of the galaxy, considering the approaching and receding regions, to ensure the reliability of the kinematic properties.
Since our sample is composed only of high-mass galaxies ($10^{10.6}-10^{11.1}$~$\textrm{M}_\odot$), we neglect any asymmetric drift correction and assume that $V_{rot}(r)\equiv V_{cir}(r)$.
Upon running $^{\rm{\rm{3D}}}$\textsc{Barolo}, we generate maps of intensity and velocity for each HI data cube. 
An example of the velocity maps and their respective models used to derive the RCs of our sample are shown in Figure~\ref{fig:plot_Barolo}.
We note that this analysis is consistent with previous work on the VIVA galaxies \citep[e.g.,][]{2021ApJ...916...26R}; we have reproduced the analysis here for all of the galaxies in our sample in the interest of homogeneity.

In order to mitigate the presence of outlier data points in the RCs, we implement an exclusion criterion based on deviations from the mean value. Specifically, we exclude any data point in the RC that deviates by more than 30\% from the mean value computed using the three nearest neighboring points. 
This approach helps to ensure a more reliable and robust representation of the RCs by removing potentially anomalous measurements.
The efficacy of this criterion is particularly apparent in the case of the MW analog NGC~2903, wherein the lack of data at a radial distance of  r $\sim$~8-9~kpc from the galaxy centre (visible in Figure~\ref{fig:plot_Barolo}) resulted in inaccurate velocity data points in this region.

\subsection{Rotation curve decomposition and mass models} \label{subsec:mass_models}
To investigate how DM is distributed in our MW-analog systems we decompose each RC into the individual mass contributions (stars, gas, and DM halo).
An RC is usually represented by the quadrature sum of the matter components of a galaxy:
\begin{equation}\label{eq:vc}
     v_{\rm{c,total}}^2(R)= -r \frac{\partial \Phi(R)}{\partial R} =
     v_{\rm{c,\star}}^2(R) 
     + v_{\rm{c,gas}}^2(R) 
     + v_{\rm{c,halo}}^2(R)
\end{equation}
where $\Phi(R)$ is the gravitational potential, and $v_{c,\star}$, $v_{c,\rm{gas}}$ and $v_{c,\rm{halo}}$ are the circular velocity due to the stellar and gaseous discs and the DM halo components, respectively. 
We perform RC decompositions assuming three mass density profiles, discussed here separately.

\subsubsection{Dark matter halo}\label{subsec:nfw}
Throughout this work we consider two radial profiles of DM distribution within the context of a cold-DM model to analyze the DM halo contribution within our MW analogs. The Navarro-Frenk-White profile \citep[hereafter NFW;][]{1996ApJ...462..563N,1997ApJ...490..493N} is based on N-body simulations, according to a flat $\Lambda$CDM cosmology and is used to describe the behavior of DM particles in halos over a wide range of mass scales, from dwarf galaxy halos to galaxy clusters.
The NFW density profile follows the shape:
\begin{equation}\label{eq:nfw}
    \rho_{\rm NFW}(r) = \frac{\rho_{\rm crit}\,\delta_c}{\frac{r}{r_s}\,\left(1+ \frac{r}{r_s}\right)^2}
\end{equation}
where $\rho_{\rm NFW}(r)$ is the DM density at radius $r$; the critical density is $\rho_{\rm crit}=3H_0^2/8\pi G$, where $H_0$ is the current value of Hubble's constant; $r_s=r_{200}/c$ is the scale radius, where c is refereed to as the concentration parameter of the halo, which is related to the characteristic overdensity, $\delta_c$, through the relation
\begin{equation}\label{eq:delta}
    \delta_c=\frac{200}{3}\,\frac{c^3}{\ln(1+c)-\frac{c}{1+c}}.
\end{equation}
The mass of the DM halo, $\textrm{M}_{200}$, can be derived through $r_{200}$ via $\textrm{M}_{200}=~200\rho_{crit}(4\pi/3)r_{200}^3$, where the
average density is 200 times the critical density of the universe.

We also use the Einasto profile \citep[][]{Einasto} to study the DM component. Just as for the NFW profile, the Einasto assumes a flat $\Lambda$CDM cosmology. Its density profile is given by
\begin{equation}
\rho_{Ein}= \rho_s\, \exp\left\{-\frac{2}{\alpha_\epsilon}\left[\left(\frac{r}{r_s}\right)^{\alpha_\epsilon} -1\right]\right\},
\end{equation}
which introduces an additional shape parameter, $\alpha_\epsilon$. This shape parameter is related to the DM halo mass \citep[][]{2014MNRAS.441.3359D},
\begin{equation}
    \alpha_\epsilon = 0.0095\nu^2 + 0.155,
\end{equation}
where $\log_{10}(\nu)=-0.11+0.146m+0.0138m^2+0.00123m^3$ and $m=\log_{10}(M_{halo}/10^{12} h^{-1}~M_\odot)$. In this context, when $\alpha_\epsilon > 0$, the profile has a finite central density.

Although throughout this work we only consider profiles within the cold-DM model, we note that other models incorporating different types of DM (e.g., warm- or hot-DM) could impact the results we present here, and this should be regarded as a caveat of this work. 
However, \citet[][]{2022MNRAS.516.2389V} compared simulations of MW-mass galaxies with cold-DM and self-interacting DM (SIDM) models. In the presence of baryonic feedback effects, the authors found that collisional SIDM models do not produce the large differences in the inner structure of MW-mass galaxies predicted by SIDM models without baryons.

\subsubsection{Gaseous disk}\label{subsec:gas}
The gas distribution in the galactic disk plays an important role in the overall baryonic matter content of a galaxy.
In this study, we investigate the contribution of the gas component to the circular velocity by deriving the surface mass density ($\Sigma_{\rm gas}$) profile of the gas using $^{\rm{\rm{3D}}}$\textsc{Barolo}.
The gravitational potential can be written as
\begin{equation}\label{eq:potential}
\Phi_{\text{gas}}(R,\theta) = -8 G \int_0^{r_{\text{max}}} R' \Sigma_{\text{gas}}(R') \frac{K\left(\frac{4 R R'}{\sqrt{R^2 + R'^2 + 2 R R'}}\right)}{\sqrt{R^2 + R'^2 + 2 R R'}} \, dR'
\end{equation}
where $K$ is the complete elliptic integral of the first kind \citep[e.g.,][]{1968hmfw.book.....A}, given as follows:
\begin{equation}\label{eq:elliptic}
    K(m)=\int_{0}^{\pi/2} \frac{d\theta}{\sqrt{1-m~\sin^2{\theta}}}
\end{equation}
Solving equation (\ref{eq:elliptic}) numerically, we take the derivative of equation (\ref{eq:potential}) to calculate the circular velocity through equation (\ref{eq:vc}). Note that the integral in equation~(\ref{eq:potential}) is taken over the whole galaxy, where upper limit $r_{max}$ represents the faintest radius of HI data.

In order to take into account the helium contribution to the gaseous disk, we assume that $\langle \Sigma_{\textrm{gas}}\rangle=1.36\times\langle \Sigma_{\textrm{HI}}\rangle$, where the factor 1.36 represents the correction by the helium mass contribution \citep[e.g.,][]{2022ApJ...935L...5C}. 
Since molecular gas is generally a minor, subdominant component compared to stars and neutral hydrogen \citep[e.g.,][]{2003ApJ...596..957S,2016AJ....151...94F,2016AJ....152..157L}, especially at the galactocentric distances of interest in this work (corresponding to the Sun’s location in the MW), we here neglect the contribution from molecular gas.

\subsubsection{Stellar disk}
To account for the stellar mass 
component we rely on S$^4$G imaging\footnote{Masks, general photometric parameters and surface brightness profiles can be accessed at: \url{https://irsa.ipac.caltech.edu/data/SPITZER/S4G/galaxies/}}. For this, we calculate the emission at 3.6 and 4.5~$\mu$m at several circular concentric annuli to disentangle the contributions from each band for each radius along the galaxy's semi-major axis. We then apply the relation introduced by \citet[][]{2015ApJS..219....5Q} to calculate the stellar mass (equation \ref{eq:stellar_mass}) through the fluxes in 3.6 and 4.5~$\mu$m.
This method for deriving stellar mass is particularly useful in order to counter problems due to the disk-halo degeneracy \citep[e.g.,][]{2018MNRAS.476.1909A,2021MNRAS.500.3579A}.
Through the constructed stellar mass maps, we derive the mass surface density ($\Sigma_\star$) at each radius. 
Our analysis returns very similar stellar contributions compared to previous studies without any correction by inclination, such as for the \textit{Spitzer Photometry and Accurate Rotation Curves} \citep[SPARC;][]{2016AJ....152..157L}.
With this approach we eliminate the need of assuming a fixed mass-to-light ratio, which is one of the sources of mass uncertainties \citep[e.g.,][]{2008AJ....136.2648D}.
Once we have $\Sigma_\star$ of the stellar component, we perform the same analysis presented in $\S$\ref{subsec:gas} for the gaseous disk.

After deriving the contributions from the gaseous and stellar disks, we can proceed with our RC decomposition to isolate the DM halo contribution.
Figure~\ref{fig:corner_plot} displays the final RC decomposition for the MW analogs in our sample, with the DM halo curve representing what remains of the total fitted curve after considering the contributions from gas and stars.

\subsection{MCMC-based derivation of the DM halo mass and concentration}\label{subsec:MCMC}
In this work we perform a Markov Chain Monte Carlo (MCMC) analysis to estimate the DM halo mass, $M_{200}$, and concentration parameter, $c$, (hereafter $\textbf{x}=(M_{200},c)$) for the RC of each galaxy in our sample. 
For this analysis, we assume that all the data points in the RC are independent. 
To inform our analysis, we implement priors of $9.0 \le \log_{10}(M_{200}) \le 13.0$ and $1 < c < 50$.

The MCMC algorithm generates a chain of samples from a probability density function (PDF), which represents the joint posterior distribution of the model parameters. We compute the posterior distributions of these parameters with a Bayesian approach, where we consider a standard likelihood function $\chi^2$ for the data $v_{cir}$ as given by
\begin{equation}
    \chi^2= -\ln \mathcal{P}(v_{cir}|v_{mod}(\textbf{x}))=\sum_{i=1}^N~\frac{1}{2}\left[\frac{v_{cir}(r_i) - v_{mod}(r_i)}{err_{v_{cir}}} \right]^2
\end{equation}
where $v_{cir}$ and $v_{mod}$ are the observed and model circular velocity at radius $i$-th, respectively, and $err_{v_{cir}}$ is the observed circular velocity error at the corresponding radius. From these distributions, we calculate the 16th, 50th, and 84th percentiles to define the final set of values and their uncertainties. 
We report the resulting values in Table~\ref{tab:percentiles}.

Here we use an MCMC routine based on the \sc{Python} \normalfont package \sc{emcee} \normalfont \citep{2013PASP..125..306F}, using 50 walkers and 30~000 iterations for each run.
In order to ensure convergence of the MCMC algorithm, we discard the first 25\% of the chains as burn-in samples \citep[e.g.,][]{2022MNRAS.514.3329M}.
To compare our results, we run our MCMC code for both NFW and Einasto profiles in order to find the best-fit curve to the observational data and the decomposed contribution from the DM halo.

To visualize the correlations between the fitted parameters we generate correlation plots, as shown in the right panels of the Figure~\ref{fig:corner_plot} for all galaxies in our sample. The correlation plot displays the one- and two-dimensional marginalized posterior probability distributions for each parameter (in this case, $M_{200}$ and $c$), allowing us to see how the uncertainties in one parameter affect the uncertainties in the other one. 
We highlight the good convergence found for all MW analogs as we show in the corner plots of Figure~\ref{fig:corner_plot}, where the probability distribution function closely resembles a normal gaussian distribution, exhibiting symmetric distributions on both sides around the mean value, indicated by the orange vertical line in the individual figures of the corner plot.

%
%
%
%
\section{Results and Discussion}
\label{sec:results}
Once we have constructed and decomposed the RCs of our MW-analog sample into their stellar, gas and DM components, we are in a position to explore the baryonic and DM content and distribution within these systems. 
Figure \ref{fig:corner_plot} displays the main results of our analysis, showing the stellar emission in 3.6~$\mu$m with overlaid HI emission contours for all galaxies in our sample, as well as the decomposition of its RC.

\subsection{Dark matter distribution in MW analogs}\label{subsec:DM dist}
\begin{figure*}[t]
  \centering
\includegraphics[width=0.93\textwidth]
{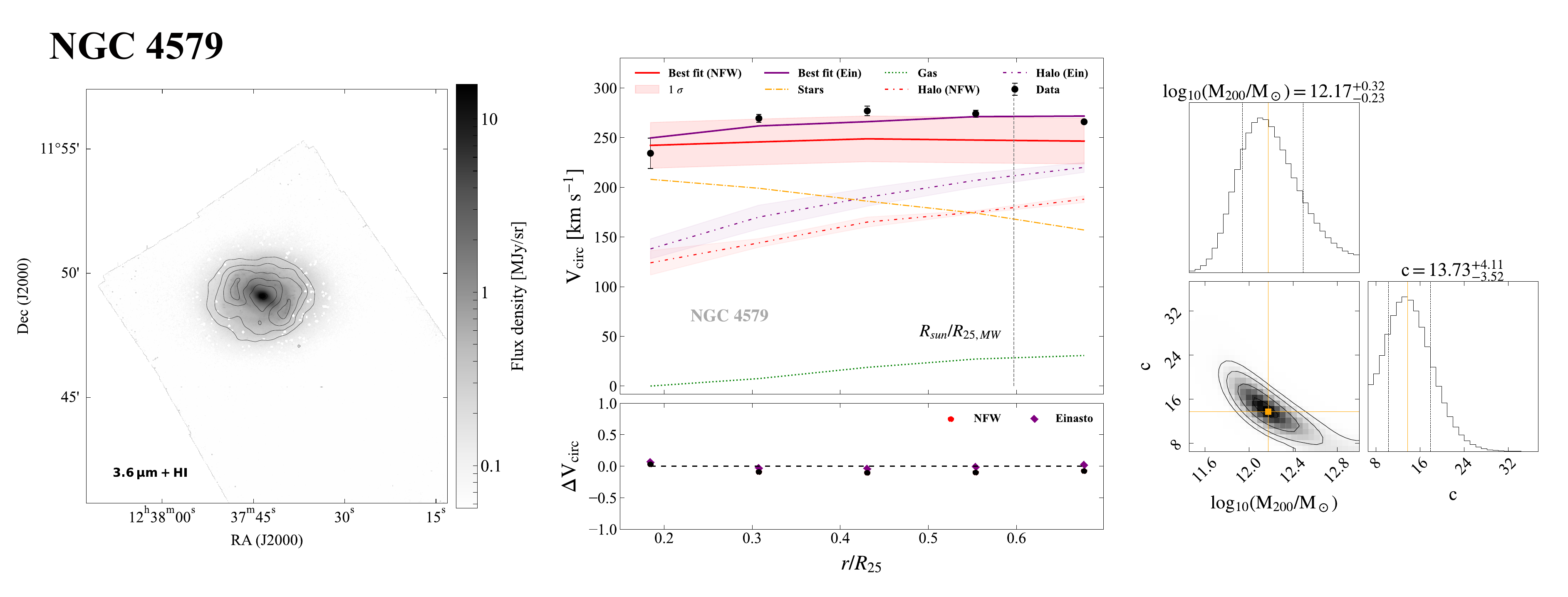}
    \caption{
    \textit{Left column:} Mid-IR emission at 3.6~$\mu$m \citep[background image; from S$^4$G;][]{2010PASP..122.1397S} for each MW analog in our sample, with overlaid contours corresponding to the HI intensity maps. 
    In all mid-IR images we employ masks provided by \citet{2010PASP..122.1397S} to mitigate contamination from foreground stars.
    \textit{Middle column:} Rotation curve decomposition into the different mass contributions from the gas (dotted green curve), stars (dashed orange curve), and DM halo (dot-dashed purple line). For the DM halo contribution we show two different curves. The purple one corresponds to the DM contribution based on the Einasto profile, while the red one is based on the NFW profile. The shadowed region around the halo's decomposed curves represents the $1\sigma$ of the distribution of values at each radius, generated from a thousand synthetic spectra following a bootstrap-based approach. The velocities derived from the HI data with their respective error bars are shown as full black circles. The solid red line represents the best-fit curve to the observational data based on the NFW profile, along with the corresponding 1$\sigma$ confidence level. We also show as the purple solid line the best-fit curve based on the Einasto profile. For all galaxies, the vertical dashed black line represents the radial location r/R$_{25}=0.6$, corresponding to the radial position of the Sun within the MW disk.
    The bottom panel in the middle column shows the residuals of the fit to the HI data.
    \textit{Right~column:} Correlation plot illustrating the MCMC analysis results based on the NFW profile. Solid orange line denotes the 50th percentile value, while the dashed black lines are the 16th and 84th percentiles, adopted as uncertainties on the values reported in Table~\ref{tab:percentiles}.}
    \label{fig:corner_plot}
\end{figure*}

As part of our analysis, we decompose the RCs of our MW analogs into their different stellar, gas and DM mass contributions ($\S$\ref{subsec:mass_models}). 
Figure~\ref{fig:corner_plot} shows the RC for each system, with this and the best-fitting curve as determined by the MCMC analysis described in $\S$\ref{subsec:MCMC}.
Values of the $\chi^2/N$ for all RC best-fitting based on both NFW and Einasto profiles are presented in Table~\ref{tab:percentiles}. Figure~\ref{fig:corner_plot} also shows the corresponding residuals between the data and the best-fit curve.
\begin{figure*}[t]
  \centering
\includegraphics[width=0.93\textwidth]{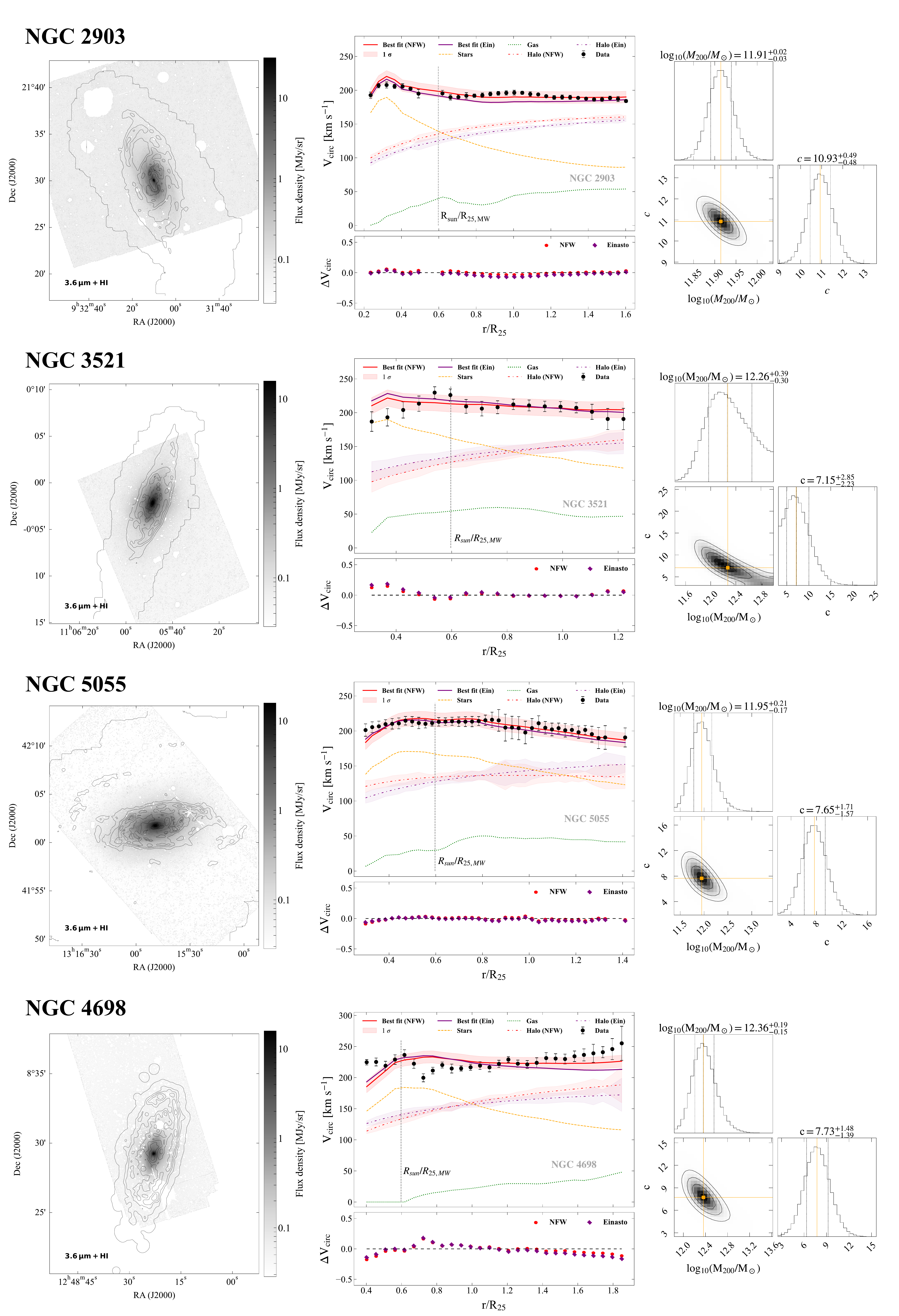}
\end{figure*}

We are particularly interested in using these MW analogs as external references to the DM matter distribution of our own galaxy. 
In particular, we seek to estimate the value of the DM density at the galactocentric distance that corresponds to the Sun’s position within the MW.
To establish the corresponding position of the Solar neighborhood in each MW analog, we first consider expressing the Sun’s location within the MW in units of R$_{25}$\footnote{R$_{25}$ is defined as the radius corresponding to the 25~mag~arcsec$^{-2}$ isophote in the B-band. The R$_{25}$ values of our MW analogs come from the HyperLeda database.}, a parameter that is specific to each galaxy. 
For the MW, we adopt R$_{25}$ as 13.4 kpc \citep[][]{1998Obs...118..201G}, with the Sun’s galactocentric distance set at 8.15$~\pm$~0.15~kpc and orbital velocity of 236~$\pm$~7~km~s$^{-1}$ \citep{2019ApJ...885..131R}. 
The Solar neighborhood is thus located at r/R$_{25}$~=~0.60~$\pm$~0.01 from the center of the MW.
We proceed to express the RCs for our MW-analog sample in distance units that, rather than expressed in physical units (kpc), are normalized by each galaxy’s R$_{25}$ value. 
The radial distance at which the Sun’s location corresponds in each MW analog of our sample (r/R$_{25}\sim0.6$) is highlighted in Figure~\ref{fig:corner_plot} by the dashed vertical line (middle panels).

For all MW analogs, we find comparable contributions originating from the baryonic and DM components at the r/R$_{25}\sim$0.6 galactocentric distance. 
We note that in the case of NGC~3521 and NGC~4579, the best-fit curve seems to overestimate the circular velocity in the very innermost region. 
This can be solved by separating the stellar component into its contribution from both the bulge and disk.
However, given the galactocentric region of interest in this work, a separate bulge and disk contribution would have no significant impact on our final results. 
This also applies to the molecular component, as its contribution is not expected to be significant in these regions (see $\S$\ref{subsec:gas}).

\begin{table}
\begin{center}
\caption{MCMC analysis results}
\begin{center}
\begin{tabular}{ ccccc }
\hline
    Galaxy & $\log_{10}\left(\frac{M_{200}}{M_\odot}\right)_{-\sigma}^{+\sigma}$ & $c_{-\sigma}^{+\sigma}$ & $\frac{\rho_{dm}}{\textrm{GeV~cm$^{-3}$}}$ & $\chi^2/N$\\
    (1) & (2) & (3) & (4) & (5)\\
\hline
\hline
\multicolumn{5}{c}{NFW-based approach}\\
    NGC 2903 & $11.91_{-0.03}^{+0.02}$ & $10.93_{-0.48}^{+0.49}$ & $0.23$ & 1.43\\
    NGC 3521 & $12.27_{-0.30}^{+0.39}$ & $8.43_{-2.50}^{+2.95}$ & $0.27$ & 0.75 \\
    NGC 4579 & $12.17_{-0.23}^{+0.33}$ & $13.80_{-3.60}^{+4.17}$ & $0.46$ & 2.34\\
    NGC 4698 & $12.36_{-0.15}^{+0.19}$ & $7.73_{-1.39}^{+1.48}$ & $0.38$ & 1.27 \\
    NGC 5055 & $11.95_{-0.17}^{+0.21}$ & $7.65_{-1.57}^{+1.71}$ & $0.21$ & 1.35 \\
    
\multicolumn{5}{c}{Einasto-based approach}\\
    NGC 2903 & $12.04_{-0.97}^{+1.04}$ & $8.26_{-0.07}^{+0.08}$ & $0.24$ & 3.40\\
    NGC 3521 & $11.99_{-1.00}^{+1.01}$ & $8.31_{-0.44}^{+0.46}$ & $0.31$ & 1.73 \\
    NGC 4579 & $12.36_{-1.00}^{+1.03}$ & $7.61_{-0.53}^{+0.52}$ & $0.55$ & 2.72\\
    NGC 4698 & $11.78_{-0.98}^{+1.00}$ & $9.02_{-0.26}^{+0.27}$ & $0.29$ & 5.90 \\
    NGC 5055 & $11.94_{-0.89}^{+0.81}$ & $7.43_{-0.28}^{+0.28}$ & $0.32$ & 0.80 \\

\hline
\end{tabular}
\end{center}
\label{tab:percentiles}
\end{center}
\footnotesize \bf{Notes:} \normalfont (1)~Galaxy identifier; (2)~M$_{200}$ and (3)~concentration parameter for each MW-analog galaxy derived from our MCMC-analysis (see $\S$\ref{subsec:MCMC}); (4)~DM density at r$\sim$0.6~R$_{25}$ (the corresponding location of the solar neighbourhood) in each MW analog; (5)~$\chi^2$ per degree of freedom returned by the fit. Lower and upper uncertainties correspond to the 16th and 84th percentiles found based on both NFW and Einasto profiles.
\end{table}

For all our RC decompositions, we note that the NFW profile fits better to our HI observational data, especially in the outskirts.
We verify that the best fit curves based on Einasto present slightly higher values of $\rho_{DM}$ in the innermost regions when compared to the NFW best fit curves (except for NGC~4698).
This is well explained given the Einasto's characteristic of having a flatter inner region and more gradual density increase towards the center than the NFW profile. Based on our analysis, we estimate that the radius at which the contribution of DM becomes equal to that of the baryonic matter in MW-analog galaxies occurs around $r$~$\sim$~0.65~$\pm$~0.10~$R_{25}$.
We note that this result still holds regardless of the assumed DM profile (NFW or Einasto).
This outcome suggests that DM begins to dominate over baryonic matter within a regime where the luminous contribution from stars remains relatively high.
We also estimate that the typical contributions from baryonic (stars + gas) and DM around the corresponding location of the solar neighbourhood are quite comparable, with baryons and DM corresponding to 55\% and 45\% of the total matter, respectively.
This result could be linked to the spherical symmetry we assume in our analysis for the DM component, although recent findings indicate that the shape of the MW's DM halo is indeed very close to spherical \citep[e.g.,][]{2023MNRAS.524.2124P}.

We note that NGC~4579 and NGC~4698, both well within the selection criteria that resulted in our MW analog sample, have some peculiarities that are worth mentioning.
In the case of NGC~4579, the HI extension is quite limited, not reaching R$_{25}$.
This is most likely due to the moderately HI-deficiency ($def_{HI}\sim1$), as confirmed by \citet[][]{2009AJ....138.1741C}.
For MW analog NGC~4698, a short tail signature in the HI (see Figure~\ref{fig:plot_Barolo}) may indicate a recent history of a minor merger, also suggested by \citet[][]{2009AJ....138.1741C}.
We also note that these galaxies have been identified as hosting AGN activity, both of them being classified as LINER galaxies \citep[][]{2000ApJ...532..323P,2003ApJ...594..704G}.
However, given that these galaxies fit well within our sample selection criteria ($\S$\ref{subsec:sample criteria}) and that their RCs, DM halo masses, nor concentration parameters stand as outliers, we do not exclude them from our study.

\subsection{A local dark matter density for the MW, based on MW analogs}\label{subsec:LDMD}
\begin{figure}[h!]
\centering
\includegraphics[width=0.53\textwidth]{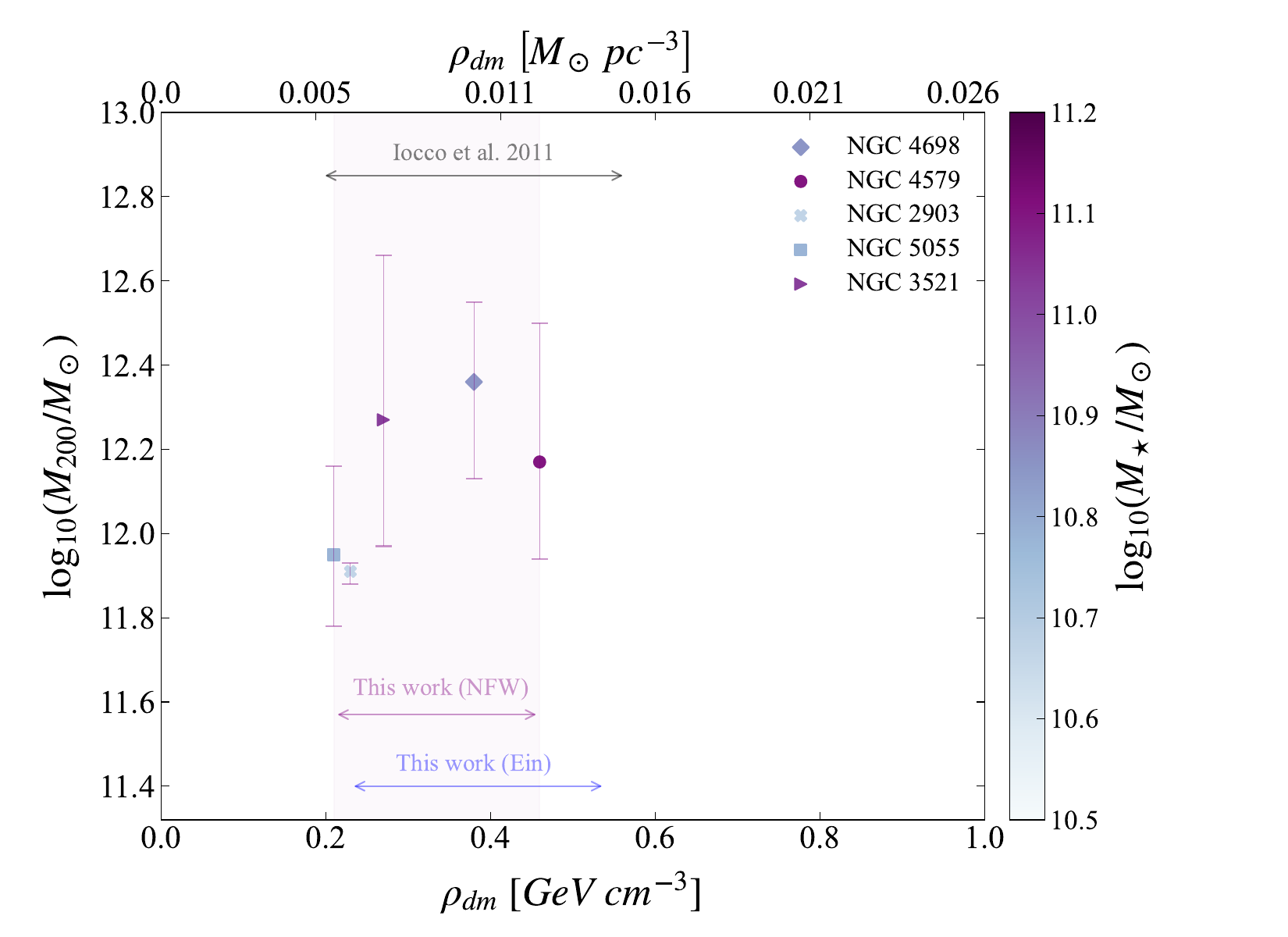}
    \caption{DM density estimates at the corresponding Sun’s location (r/R$_{25}$=0.6) for our MW-analog sample. The error bars correspond to the 16th and 84th percentiles of the probability distribution function. Galaxies are colour-coded by their stellar mass. Double-headed arrows show our suggested window of values based on the NFW profile (purple) and Einasto profile (blue). We show, for reference, the interval of values for the MW local DM density by \citet[][]{2011JCAP...11..029I}.}
    \label{fig:LDMD}
\end{figure}

We are entering an era characterized by experiments of ultra-high precision in the direct detection of DM (e.g., LZ, \citealt{2015arXiv150902910T}; XENON, \citealt[][]{2017PhRvL.119r1301A}). 
Such experiments require accurate values of the local DM density in order to estimate the number of detectable particles
\citep[e.g.,][]{1996APh.....6...87L,2020PhRvD.101e2002A}.
Recent studies on the local MW DM density have converged to values around 0.3~GeV~cm$^{-3}$\footnote{It is most common seeing local DM density values in GeV~cm$^{-3}$.
An useful conversion is 0.008 M$_\odot$~pc$^{-3}$ = 0.3 GeV~cm$^{-3}$. We present both forms in Figure~\ref{fig:LDMD}.}, as we show in the literature compilation presented in Table \ref{tab:LDMDe_estimates}.

\begin{figure}
\centering
\includegraphics[width=0.5\textwidth]{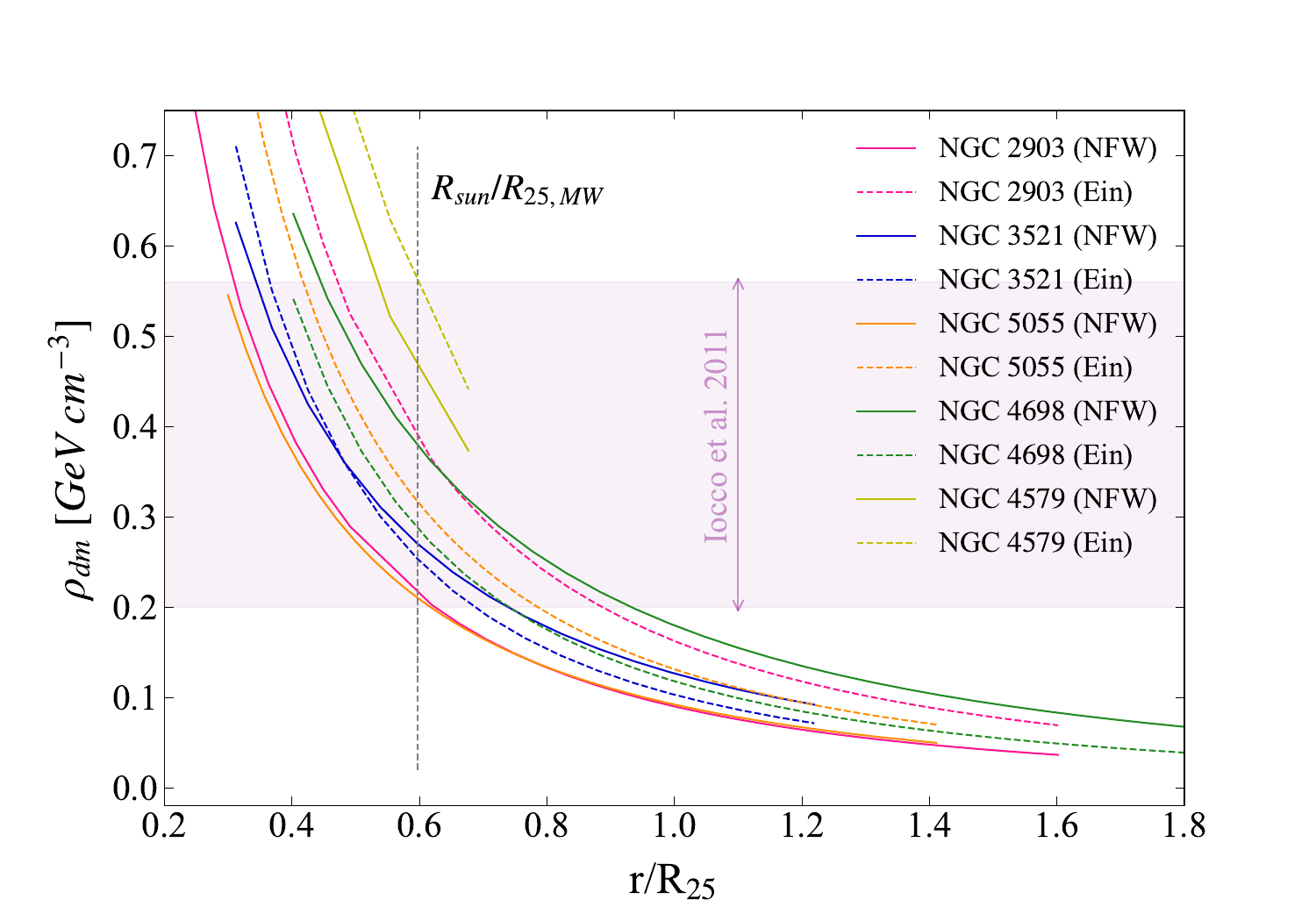}
    \caption{Radial profile of the DM density of our MW analogs. 
    All of our galaxies show a prominent decreasing trend of $\rho_{dm}$ as a function of distance, consistent with NFW (solid curves) and Einasto (dashed curves) halos.
    We highlight the Sun corresponding position as a vertical gray dashed line. 
    The HI profile of NGC~4579 does not extend out to large radii, most likely due to its moderately HI-deficiency of $\sim1$ \citep[][]{2009AJ....138.1741C}.
    We also show the values for the MW local DM density \citep[][]{2011JCAP...11..029I}. As in previous figures, we normalize the x-axis by the galaxy’s R$_{25}$.}
    \label{fig:rhos}
\end{figure}

Although two main kinematic-based approaches have been used to calculate the local DM density -- measuring vertical motion of stars in the solar neighbourhood \citep[e.g.,][]{1922ApJ....55..302K,1922MNRAS..82..122J,1932BAN.....6..249O,1987A&A...180...94B,1998A&A...329..920C,2004MNRAS.352..440H,2012ApJ...756...89B,2013ApJ...772..108Z} and assuming a global shape for the MW DM halo -- other techniques have also been used.
For instance, in the recent study by \citet{2023MNRAS.524.2124P} the authors used Galactic globular clusters to explore constraints on the MW's DM halo, in particular its shape, and propose a value for the local DM density, consistent with the other works.
We also note recent work by \citet[][]{Chakrabarti+21}, where the authors used extreme-precision time-series measurements to constrain the Galactic potential, and calculate the local DM density. Currently, this technique is the most direct measurement possible, although the lack of a large number of pulsars turns this technique not as precise as some of the methods mentioned above.
In Figure \ref{fig:rhos} we examine the shapes of the DM density curves by displaying the radial density profile of DM for the galaxies within our sample. For all MW analogs, we observe a sharp decrease in $\rho_{dm}$ with the galactocentric distance, as we expect for typical NFW and Einasto halos.
Although the pronounced radial decrease is similar for both NFW and Einasto, we note, however, that the curves based on Einasto present slightly higher values of $\rho_{DM}$ in the innermost regions when compared to the curves based on NFW (except for NGC~4698).
We measure the DM density at r/R$_{25}\sim0.6$ for each galaxy in our sample using both NFW and Einasto profiles, with the objective of building a window of values for the DM density at the radial distance at which the Sun’s location corresponds in each MW analog of our sample. With values in the range of 0.21--0.55~GeV~cm$^{-3}$, our estimates for the local DM density are in agreement with those from previous works in the literature with a focus on MW (see Table~\ref{tab:LDMDe_estimates}). We show these values in Figure~\ref{fig:LDMD} as a function of the host DM halo masses (assuming M$_{200}=\,$M$_{vir}$).  We also display in Figure~\ref{fig:LDMD} the range of values suggested by \citet{2011JCAP...11..029I} based on a best-fit to the combination of microlensing and the MW's RC data, which entirely encompasses our (narrower) proposed interval. 

With our work, based on the DM distribution in MW analogs assuming a NFW profile, we propose a novel window for the DM density in the solar neighborhood, covering the range of 0.21--0.55~GeV~cm$^{-3}$. When considering both NFW and Einasto profiles, our range of values covers the interval from 0.21 to 0.55~GeV~cm$^{-3}$, as we show in Figure~\ref{fig:LDMD}. Our estimates find support in the recent result by \citet[][]{2023arXiv230316217P}, who used a sample of 198 simulated MW/M31 analogs with Illustris TNG-50 simulation to estimate the DM density in the solar neighbourhood (at distances of 7--9~kpc from the galaxy’s centre).

\begin{table}
\begin{center}
\caption{A summary of previous values for the  DM density at the Sun’s location for the MW from past literature, and our mean value based on our MW-analog sample, for the corresponding location of the Sun at r/R$_{25}=0.6$}

\begin{tabular}{ cccc }
\hline
    Work & $\rho_{dm}$     & $V_0$\\
         & [GeV~cm$^{-3}$] & [km~s$^{-1}$]\\
    (1)  & (2)             & (4) \\
\hline
\hline
    \citet[][]{2011JCAP...11..029I} & 0.20 -- 0.56 & 240 \\
    \citet[][]{2012ApJ...756...89B} & 0.30 $\pm$ 0.10 & 220 \\
    \citet[][]{2013PASJ...65..118S} & 0.40 $\pm$ 0.04 & 238 \\
    \citet[][]{2014MNRAS.445.3133P} & 0.59 & 240 \\
    \citet[][]{2015JCAP...12..001P} & 0.42 $\pm$ 0.25 & 230 \\
    \citet[][]{2016MNRAS.463.2623H} & 0.32 $\pm$ 0.02 & 240\\
    \citet[][]{2017MNRAS.465...76M} & 0.38 $\pm$ 0.04 & 233\\
    \citet[][]{2019MNRAS.487.5679L} & 0.58 $\pm$ 0.09 & 240\\
    \citet[][]{2019ApJ...871..120E} & 0.30 $\pm$ 0.03 & 229 \\
    \citet[][]{2019JCAP...10..037D} & 0.30 -- 0.40  & 229\\
    \citet[][]{2019JCAP...09..046K} & 0.43 $\pm$ 0.03 & 240\\
    \citet[][]{2020MNRAS.494.4291C} & 0.34 $\pm$ 0.02 & 229 \\
    \citet[][]{2020Galax...8...37S} & 0.36 $\pm$ 0.02 & 238 \\
    \citet[][]{Chakrabarti+21} & 0.13 $\pm$ 1.31 & -\\
    \citet[][]{2023MNRAS.524.2124P} & 0.23 $\pm$ 0.02 & - \\
    This work & 0.21 -- 0.55 & 236 \\
\hline
\end{tabular}
\label{tab:LDMDe_estimates}
\end{center}
\footnotesize \bf{Notes:} \normalfont (1)~Reference; (2)~Local DM density reported; (3)~Solar galactocentric radius assumed in the analysis; (4)~Solar radial velocity assumed in the analysis. Observation: \citet[][]{2023MNRAS.524.2124P} assumed $V_0$ as a free parameter of their modelling. \citet[][]{Chakrabarti+21} used pulsar timing to directly measure the Galactic acceleration and estimate their own local DM density value.
\end{table}

\subsection{Stellar-to-halo mass relation}\label{subsec:SHMR}
The stellar-to-halo mass relation (SHMR) has been extensively used to better understand the star formation efficiency as a function of the host DM halo mass, providing valuable insights into galaxy formation models \citep[e.g.,][]{2013MNRAS.428.3121M,2019A&A...626A..56P,2020MNRAS.499.5656R,2020A&A...634A.135G,2021A&A...649A.119P,2022MNRAS.514.3329M,2022A&A...664A..61S}.
These studies provide relatively consistent formats of the SHMR curve, pointing to a peak of galaxy formation efficiency for DM halo masses of $\sim10^{12.0}$~M$_{\odot}$, coincident with the DM halo masses for the MW and M31 \citep[e.g.,][]{2019A&A...626A..56P, 2010A&A...511A..89C}.
In higher-mass halos, feedback processes attributed to the presence of an AGN are expected to be responsible for driving down this efficiency, heating the halo gas and hindering star formation \citep[e.g.,][]{1998A&A...331L...1S, 2006MNRAS.365...11C}. In low-mass halos, this is likely due to feedback from high-mass stars, which generate strong winds that eject gas, also impacting the gas available for star formation \citep[e.g.,][]{2012MNRAS.421.3522H}.

Recent studies have consistently indicated low star formation efficiencies, with the stellar-to-baryonic mass ratio peaking within the range of 20--30\% \citep[e.g.,][]{2018AstL...44....8K, 2019A&A...626A..56P, 2021NatAs...5.1069C}.
Furthermore, evidence for a bimodality in the SHMR has drawn attention, with the suggestion that red galaxies tend to inhabit higher-mass halos, while blue ones exhibit a preference for less massive ones \citep[e.g.,][]{2021NatAs...5.1069C}.

\begin{figure}
\centering
\includegraphics[width=0.48\textwidth]{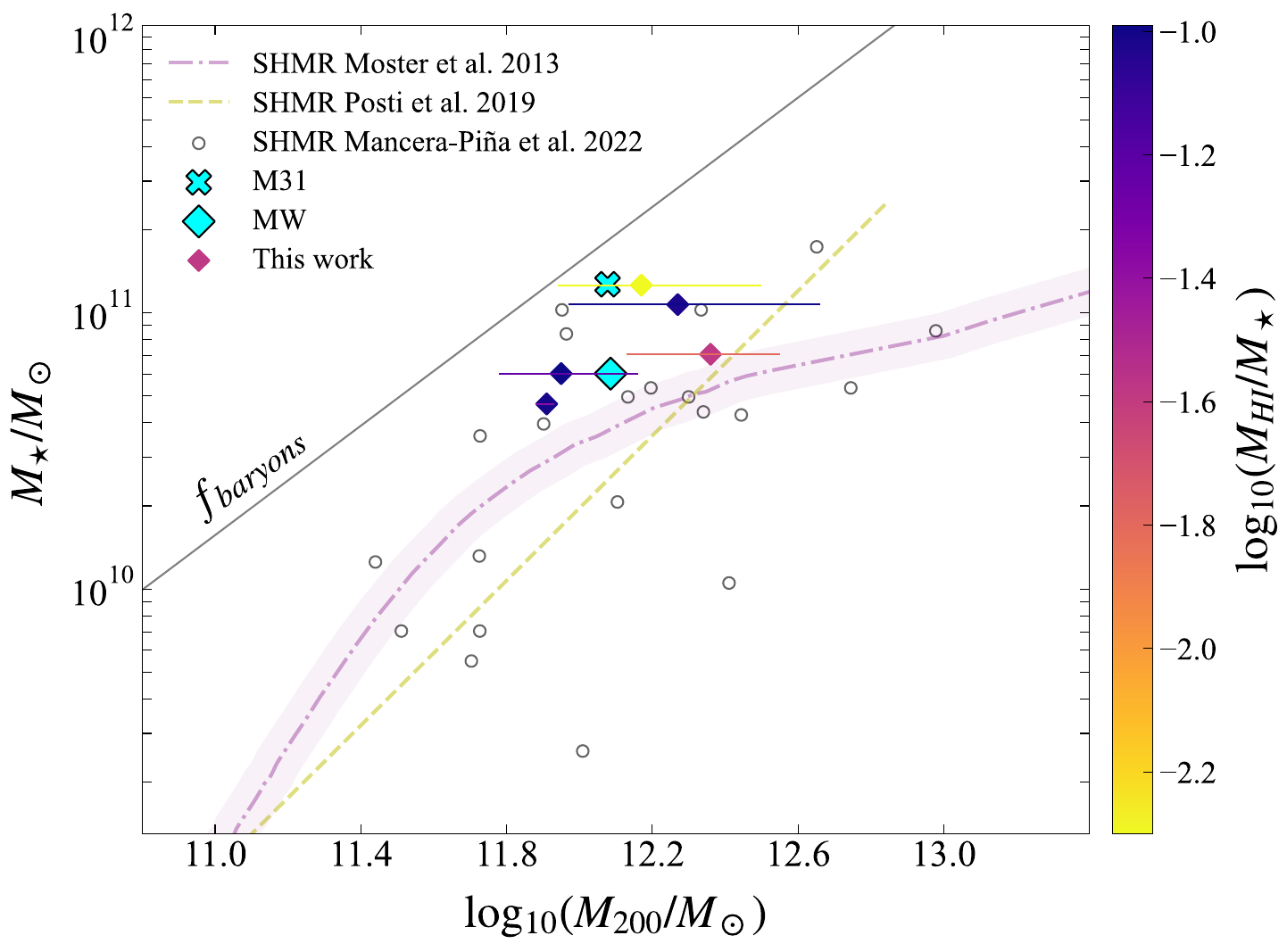}
    \caption{Stellar-to-halo mass relation of our sample of MW analogs. Our galaxies are marked as diamonds, colour-coded by their stellar-to-HI gas mass ratio.
    The MW is shown as a filled cyan diamond \citep[][]{2019A&A...626A..56P}, while Andromeda (M31) is shown as a cyan cross \citep[][]{2010A&A...511A..89C}.
    We also show the stellar-to-halo mass relations suggested by previous works in the literature: the purple dot-dashed curve represents the relation proposed by \citet[][]{2013MNRAS.428.3121M}, whereas the yellow dashed line represents a linear relationship in the $\log\times\log$ plane as suggested by \citet[][]{2019A&A...626A..56P}.
    We also plot the results found by \citet[][]{2022MNRAS.514.3329M} for their galaxy sample with M$_{200}>10^{11}~$M$_\odot$ as open black circles.
    In all distributions we note a significant scatter. 
    The solid black line represents the cosmological fraction of baryons $f_b$, where $f_b=\Omega_b/\Omega_m$. This line represents a division between a forbidden zone, above which all gas has been transformed into stars.}
    \label{fig:SHMR}
\end{figure}

In Figure~\ref{fig:SHMR} we present the SHMR for our MW-analog sample, based on stellar masses from the mid-IR S$^4$G database and the DM halo masses resulting from our MCMC analysis ($\S$\ref{subsec:MCMC}).
We compare our results with a sub-sample of blue massive galaxies [$\log_{10}$(M$_\star/$M$_\odot)>9.5$] analyzed in \citet[][]{2022MNRAS.514.3329M} to provide a wider context to our findings. We also display in Figure~\ref{fig:SHMR} fits to the SMHR based on earlier work by \citet[][]{2013MNRAS.428.3121M} and
\citet[][]{2019A&A...626A..56P}. The former results from a multi-epoch abundance matching model based on the Millennium simulations for massive and lower mass halos \citep[][]{2005Natur.435..629S,2009MNRAS.398.1150B}. The latter analyzes a sample of 110 galaxies from the SPARC survey \citep[][]{2016AJ....152..157L}, taking into account the effects of stellar and AGN feedback. Ultimately, all of these relations are limited by the cosmological fraction of baryons, where all gas has been transformed into stars.

Our sample brings a few additional data points into this picture. The resulting distribution supports a dependence between stellar mass and host DM halo mass, with the mass range [$\log_{10}$(M$_\star/$M$_\odot)\sim$~10.5--11.0]
covered by our MW analog sample being found within DM halos of masses up to $\log_{10}$(M$_{200}/$M$_\odot)$=$12.5$.
This is consistent with the colour bimodality discussed in \citet[][]{2021NatAs...5.1069C}.
Within the stellar mass interval covered by our sample, the relation saturates somewhat, as the distribution transitions from the steeper trend for galaxies within the low-mass regime [$\log_{10}$(M$_\star/$M$_\odot)<10.5$] to the shallower trend held by galaxies at the high-mass regime [$\log_{10}$(M$_\star/$M$_\odot)>11.0$]. 
This is consistent with cosmological simulations integrating black hole growth and AGN feedback \citep[e.g., SIMBA;][]{2019MNRAS.486.2827D}.
It is also in line with works based on observational data of SDSS galaxies \citep[][]{2000AJ....120.1579Y}, where satellite galaxy kinematics were used to trace the DM halo masses of central galaxies \citep[e.g.,][]{2011MNRAS.410..210M}.

Following this bimodality, it is expected that more gas-rich systems occupy lower-mass DM halos \citep[e.g.,][]{2022MNRAS.514.3329M}, while a myriad of processes would be responsible for diminishing the HI content in galaxies occupying higher-mass DM halos \citep[e.g., AGN feedback;][]{2006MNRAS.365...11C}.
Figure~\ref{fig:SHMR} shows the HI-to-stellar mass ratio (colour bar) for our sample. 
We note that our sample not necessarily fits well with this expectation, with MW analogs displaying a diversity in HI-richness for different DM halo masses.
However, our most HI-poor galaxies (NGC~4579 and NGC~4698) indeed exhibit  higher DM halo masses compared to the rest of the sample.
The confirmed presence of an AGN in these two galaxies \citep[e.g.,][]{2000ApJ...532..323P,2003ApJ...594..704G} could explain this finding, with the AGN impacting the HI content \citep[e.g.,][]{1998A&A...331L...1S, 2006MNRAS.365...11C}.
Alternatively, a strong correlation between the DM halo concentration parameter and its mass, where high-mass DM halos typically have lower concentration parameters, could be responsible for the observed correlation between the DM halo mass and the HI content in galaxies. 
We indeed observe this general trend within our sample (see Table~\ref{tab:percentiles}), with the exception of NGC~4579. 
This is likely associated with the spatially-limited HI extension of this system, associated with it being an HI-deficient galaxy \citep[e.g.,][]{2009AJ....138.1741C}. This impacts our 
ability to constrain its concentration parameter, as can be appreciated by the comparatively higher 
uncertainties in this measurement for this galaxy.
Although the correlation between the host DM halo mass and its concentration has been an extensively investigated topic \citep[e.g.,][]{2013MNRAS.428.3121M,2019A&A...626A..56P,2020ApJS..247...31L,2022MNRAS.514.3329M}, a more in-depth exploration is beyond the scope of our current study.

%
%
%
%
\section{Summary}
\label{sec:discussion}
We present an independent approach to provide insights to the DM distribution in the MW, based on the DM distribution of a sample of 5 MW-analog galaxies. We construct rotation curves for our sample by using the three-dimensional tilted-ring kinematical modelling software $^{\rm{3D}}$\textsc{Barolo} in order to decompose them into the different mass contributions from stars, gas, and DM.
We normalize the radii of our MW analogs by their $R_{25}$ in order to find the corresponding location of the solar neighborhood in these galaxies, in the interest of drawing a parallel with the DM distribution in our own galaxy.
Our main results are:

(\textit{i}) We find that the contributions from baryonic matter (stars + gas) and DM are quite similar (55\% and 45\%, respectively) around the corresponding location of the solar neighborhood ($r\sim 0.6~R_{25}$), with a rapid shift towards a dominance of DM at larger galactocentric distances.
This result suggests that DM starts to dominate the mass contribution even in regions where stellar emission still seems to dominate.

(\textit{ii}) Using an MCMC-based approach, we calculate the mass and concentration parameter for the DM halos of our MW analogs.
Through this analysis, we estimate the DM density at the radial position corresponding to the location of the solar neighborhood in the MW.
Taking advantage of this analogy between our own galaxy and this sample of MW analogs, we provide an independently-determined window of values to be compared to the local DM density in the MW: $\rho_{dm}=0.21-0.55$~GeV~cm$^{-3}$. 
This window of values is
consistent with previous results from literature, based on more traditional approaches.

(\textit{iii}) We find that the mass range covered by our MW analog sample [$\log_{10}$(M$_\star/$M$_\odot)\sim$~10.5--11.0] points to DM halos of masses up to $\log_{10}$(M$_{200}/$M$_\odot)$=$12.5$. This result agrees with recent works that suggest a colour bimodality in the stellar-to-halo mass relation, where blue galaxies (such as our MW analogs) show a preference for DM halos in this mass range, less massive than those occupied by red galaxies.

With the emergence of projects like MeerKAT and, in the future, SKA \citep[][]{2009arXiv0910.2935B,2009pra..confE..58L}, the investigation of the HI distribution in galaxies is growing at unprecedented scales. These projects are poised to revolutionize our capacity to observe the cosmos at low- and high-redshifts, opening up new frontiers in mapping and exploring the gas content and internal kinematics of galaxies. With a continuing scientific investment in high-energy experiments to uncover the nature of the DM particle, this work puts forward the possibility of exploiting the HI exploration of MW analogs as an alternative insight to the DM distribution in our own MW.\\

%
%
%
%
\section*{Acknowledgements}
We would like to thank the anonymous referee for a productive report with comments and suggestions that improved this paper.
This work is based on observations made with the Spitzer Space Telescope, which was operated by the Jet Propulsion Laboratory, California Institute of Technology under a contract with NASA.
This research used data from VIVA survey carried out on the Karl G. Jansky Very Large Array, which is part of the National Radio Astronomy Observatory. The National Radio Astronomy Observatory is a facility of the National Science Foundation operated under cooperative agreement by Associated Universities, Inc.
This work made use of THINGS, ‘The HI Nearby Galaxy 
Survey' \citep[][]{2008AJ....136.2563W}.
We acknowledge the usage of the HyperLeda database (\url{http://leda.univ-lyon1.fr}).
NGO thanks the support of the Coordination for the Improvement of Higher Education Personnel (CAPES — the national funding agency for graduate research), the Scientific Research Honor Society (Sigma Xi; Grant ID: G20211001-465), as well as all members of the Valongo Observatory Extragalactic Astrophysics Laboratory (LASEX).
KMD thanks the support of the Serrapilheira Institute (grant Serra-1709-17357) as well as that of the Brazilian National Research Council (CNPq grant 308584/2022-8) and of the Rio de Janeiro State Research Foundation (FAPERJ grant E-26/200.952/2022), Brazil.
TSG would also like to thank the support of CNPq (Productivity in Research grant 314747/2020-6) and the FAPERJ (Young Scientist of Our State grant E-26/201.309/2021).
DCR thanks the Institute for Theoretical Physics from Heidelberg University for hospitality and support. He also acknowledges support from CNPq and \textit{Fundação de Amparo à Pesquisa e Inovação do Espírito Santo} (FAPES-Brazil).
MG acknowledges support from Rio de Janeiro State Research Foundation (FAPERJ) grant n. E-26/211.370/2021.

%
%
%
%
\section*{Data Availability}
The moment maps of the HI distribution, as well as stellar maps at 3.6 and 4.5 $\mu$m, are available on this \dataset[Zenodo repository]{https://doi.org/10.5281/zenodo.11396279}.

%
%
%
%
\bibliography{sample631}{}

\begin{thebibliography}{}
\expandafter\ifx\csname natexlab\endcsname\relax\def\natexlab#1{#1}\fi
\providecommand{\url}[1]{\href{#1}{#1}}
\providecommand{\dodoi}[1]{doi:~\href{http://doi.org/#1}{\nolinkurl{#1}}}
\providecommand{\doeprint}[1]{\href{http://ascl.net/#1}{\nolinkurl{http://ascl.net/#1}}}
\providecommand{\doarXiv}[1]{\href{https://arxiv.org/abs/#1}{\nolinkurl{https://arxiv.org/abs/#1}}}

\bibitem[{{Aalbers} {et~al.}(2022){Aalbers}, {Akerib}, {Akerlof}, {Al Musalhi}, {Alder}, {Alqahtani}, {Alsum}, {Amarasinghe}, {Ames}, {Anderson}, {Angelides}, {Ara{\'u}jo}, {Armstrong}, {Arthurs}, {Azadi}, {Bailey}, {Baker}, {Balajthy}, {Balashov}, {Bang}, {Bargemann}, {Barry}, {Barthel}, {Bauer}, {Baxter}, {Beattie}, {Belle}, {Beltrame}, {Bensinger}, {Benson}, {Bernard}, {Bhatti}, {Biekert}, {Biesiadzinski}, {Birch}, {Birrittella}, {Blockinger}, {Boast}, {Boxer}, {Bramante}, {Brew}, {Br{\'a}s}, {Buckley}, {Bugaev}, {Burdin}, {Busenitz}, {Buuck}, {Cabrita}, {Carels}, {Carlsmith}, {Carlson}, {Carmona-Benitez}, {Cascella}, {Chan}, {Chawla}, {Chen}, {Cherwinka}, {Chott}, {Cole}, {Coleman}, {Converse}, {Cottle}, {Cox}, {Craddock}, {Creaner}, {Curran}, {Currie}, {Cutter}, {Dahl}, {David}, {Davis}, {Davison}, {Delgaudio}, {Dey}, {de Viveiros}, {Dobi}, {Dobson}, {Druszkiewicz}, {Dushkin}, {Edberg}, {Edwards}, {Elnimr}, {Emmet}, {Eriksen}, {Faham}, {Fan}, {Fayer}, {Fearon}, {Fiorucci}, {Flaecher}, {Ford}, {Francis},
  {Fraser}, {Fruth}, {Gaitskell}, {Gantos}, {Garcia}, {Geffre}, {Gehman}, {Genovesi}, {Ghag}, {Gibbons}, {Gibson}, {Gilchriese}, {Gokhale}, {Gomber}, {Green}, {Greenall}, {Greenwood}, {van der Grinten}, {Gwilliam}, {Hall}, {Hans}, {Hanzel}, {Harrison}, {Hartigan-O'Connor}, {Haselschwardt}, {Hertel}, {Heuermann}, {Hjemfelt}, {Hoff}, {Holtom}, {Y-K. Hor}, {Horn}, {Huang}, {Hunt}, {Ignarra}, {Jacobsen}, {Jahangir}, {James}, {Jeffery}, {Ji}, {Johnson}, {Kaboth}, {Kamaha}, {Kamdin}, {Kasey}, {Kazkaz}, {Keefner}, {Khaitan}, {Khaleeq}, {Khazov}, {Khurana}, {Kim}, {Kocher}, {Kodroff}, {Korley}, {Korolkova}, {Kras}, {Kraus}, {Kravitz}, {Krebs}, {Kreczko}, {Krikler}, {Kudryavtsev}, {Kyre}, {Landerud}, {Leason}, {Lee}, {Lee}, {Leonard}, {Leonard}, {Lesko}, {Levy}, {Li}, {Liao}, {Liao}, {Lin}, {Lindote}, {Linehan}, {Lippincott}, {Liu}, {Liu}, {Liu}, {Loniewski}, {Lopes}, {Lopez Asamar}, {L{\'o}pez Paredes}, {Lorenzon}, {Lucero}, {Luitz}, {Lyle}, {Majewski}, {Makkinje}, {Malling}, {Manalaysay}, {Manenti}, {Mannino},
  {Marangou}, {Marzioni}, {Maupin}, {McCarthy}, {McConnell}, {McKinsey}, {McLaughlin}, {Meng}, {Migneault}, {Miller}, {Mizrachi}, {Mock}, {Monte}, {Monzani}, {Morad}, {Morales Mendoza}, {Morrison}, {Mount}, {Murdy}, {Murphy}, {Naim}, {Naylor}, {Nedlik}, {Nehrkorn}, {Nelson}, {Neves}, {Nguyen}, {Nikoleyczik}, {Nilima}, {O'Dell}, {O'Neill}, {O'Sullivan}, {Olcina}, {Olevitch}, {Oliver-Mallory}, {Orpwood}, {Pagenkopf}, {Pal}, {Palladino}, {Palmer}, {Pangilinan}, {Parveen}, {Patton}, {Pease}, {Penning}, {Pereira}, {Pereira}, {Perry}, {Pershing}, {Peterson}, {Piepke}, {Podczerwinski}, {Porzio}, {Powell}, {Preece}, {Pushkin}, {Qie}, {Ratcliff}, {Reichenbacher}, {Reichhart}, {Rhyne}, {Richards}, {Riffard}, {Rischbieter}, {Rodrigues}, {Rodriguez}, {Rose}, {Rosero}, {Rossiter}, {Rushton}, {Rutherford}, {Rynders}, {Saba}, {Santone}, {Sazzad}, {Schnee}, {Scovell}, {Seymour}, {Shaw}, {Shutt}, {Silk}, {Silva}, {Sinev}, {Skarpaas}, {Skulski}, {Smith}, {Solmaz}, {Solovov}, {Sorensen}, {Soria}, {Stancu}, {Stark}, {Stevens},
  {Stiegler}, {Stifter}, {Studley}, {Suerfu}, {Sumner}, {Sutcliffe}, {Swanson}, {Szydagis}, {Tan}, {Taylor}, {Taylor}, {Taylor}, {Temples}, {Tennyson}, {Terman}, {Thomas}, {Tiedt}, {Timalsina}, {To}, {Tom{\'a}s}, {Tong}, {Tovey}, {Tranter}, {Trask}, {Tripathi}, {Tronstad}, {Tull}, {Turner}, {Tvrznikova}, {Utku}, {Va'vra}, {Vacheret}, {Vaitkus}, {Verbus}, {Voirin}, {Waldron}, {Wang}, {Wang}, {Wang}, {Wang}, {Wang}, {Watson}, {Webb}, {White}, {White}, {White}, {White}, {Whitis}, {Williams}, {Wisniewski}, {Witherell}, {Wolfs}, {Wolfs}, {Woodford}, {Woodward}, {Worm}, {Wright}, {Xia}, {Xiang}, {Xiao}, {Xu}, {Yeh}, {Yin}, {Young}, {Zarzhitsky}, {Zuckerman}, \& {Zweig}}]{2022arXiv220703764A}
{Aalbers}, J., {Akerib}, D.~S., {Akerlof}, C.~W., {et~al.} 2022, arXiv e-prints, arXiv:2207.03764, \dodoi{10.48550/arXiv.2207.03764}

\bibitem[{{Abdullah} {et~al.}(2020){Abdullah}, {Klypin}, \& {Wilson}}]{2020ApJ...901...90A}
{Abdullah}, M.~H., {Klypin}, A., \& {Wilson}, G. 2020, \apj, 901, 90, \dodoi{10.3847/1538-4357/aba619}

\bibitem[{{Abramowitz} \& {Stegun}(1968)}]{1968hmfw.book.....A}
{Abramowitz}, M., \& {Stegun}, I.~A. 1968, {Handbook of mathematical functions with formulas, graphs and mathematical tables}

\bibitem[{{Akerib} {et~al.}(2020){Akerib}, {Akerlof}, {Alsum}, {Ara{\'u}jo}, {Arthurs}, {Bai}, {Bailey}, {Balajthy}, {Balashov}, {Bauer}, {Belle}, {Beltrame}, {Benson}, {Bernard}, {Biesiadzinski}, {Boast}, {Boxer}, {Br{\'a}s}, {Buckley}, {Bugaev}, {Burdin}, {Busenitz}, {Carels}, {Carlsmith}, {Carlson}, {Carmona-Benitez}, {Chan}, {Cherwinka}, {Cole}, {Cottle}, {Craddock}, {Currie}, {Cutter}, {Dahl}, {de Viveiros}, {Dobi}, {Dobson}, {Druszkiewicz}, {Edberg}, {Edwards}, {Fan}, {Fayer}, {Fiorucci}, {Fruth}, {Gaitskell}, {Genovesi}, {Ghag}, {Gilchriese}, {van der Grinten}, {Hall}, {Hans}, {Hanzel}, {Haselschwardt}, {Hertel}, {Hillbrand}, {Hjemfelt}, {Hoff}, {Hor}, {Huang}, {Ignarra}, {Ji}, {Kaboth}, {Kamdin}, {Keefner}, {Khaitan}, {Khazov}, {Kim}, {Kocher}, {Korolkova}, {Kraus}, {Krebs}, {Kreczko}, {Krikler}, {Kudryavtsev}, {Kyre}, {Lee}, {Lenardo}, {Leonard}, {Lesko}, {Levy}, {Li}, {Liao}, {Liao}, {Lin}, {Lindote}, {Linehan}, {Lippincott}, {Liu}, {Lopes}, {L{\'o}pez Paredes}, {Lorenzon}, {Luitz}, {Lyle},
  {Majewski}, {Manalaysay}, {Mannino}, {Maupin}, {McKinsey}, {Meng}, {Miller}, {Mock}, {Monzani}, {Morad}, {Morrison}, {Mount}, {Murphy}, {Nelson}, {Neves}, {Nikoleyczik}, {O'Sullivan}, {Olcina}, {Olevitch}, {Oliver-Mallory}, {Palladino}, {Patton}, {Pease}, {Penning}, {Piepke}, {Powell}, {Preece}, {Pushkin}, {Ratcliff}, {Reichenbacher}, {Rhyne}, {Richards}, {Rodrigues}, {Rosero}, {Rossiter}, {Saba}, {Sarychev}, {Schnee}, {Schubnell}, {Scovell}, {Shaw}, {Shutt}, {Silk}, {Silva}, {Skarpaas}, {Skulski}, {Solmaz}, {Solovov}, {Sorensen}, {Stancu}, {Stark}, {Stiegler}, {Stifter}, {Szydagis}, {Taylor}, {Taylor}, {Taylor}, {Temples}, {Terman}, {Thomas}, {Timalsina}, {To}, {Tom{\'a}s}, {Tope}, {Tripathi}, {Tull}, {Tvrznikova}, {Utku}, {Va'Vra}, {Vacheret}, {Verbus}, {Voirin}, {Waldron}, {Watson}, {Webb}, {White}, {Whitis}, {Wisniewski}, {Witherell}, {Wolfs}, {Woodward}, {Worm}, {Yeh}, {Yin}, {Young}, \& {Lux-Zeplin Collaboration}}]{2020PhRvD.101e2002A}
{Akerib}, D.~S., {Akerlof}, C.~W., {Alsum}, S.~K., {et~al.} 2020, \prd, 101, 052002, \dodoi{10.1103/PhysRevD.101.052002}

\bibitem[{{Aniyan} {et~al.}(2021){Aniyan}, {Ponomareva}, {Freeman}, {Arnaboldi}, {Gerhard}, {Coccato}, {Kuijken}, \& {Merrifield}}]{2021MNRAS.500.3579A}
{Aniyan}, S., {Ponomareva}, A.~A., {Freeman}, K.~C., {et~al.} 2021, \mnras, 500, 3579, \dodoi{10.1093/mnras/staa3106}

\bibitem[{{Aniyan} {et~al.}(2018){Aniyan}, {Freeman}, {Arnaboldi}, {Gerhard}, {Coccato}, {Fabricius}, {Kuijken}, {Merrifield}, \& {Ponomareva}}]{2018MNRAS.476.1909A}
{Aniyan}, S., {Freeman}, K.~C., {Arnaboldi}, M., {et~al.} 2018, \mnras, 476, 1909, \dodoi{10.1093/mnras/sty310}

\bibitem[{{Aprile} {et~al.}(2017){Aprile}, {Aalbers}, {Agostini}, {Alfonsi}, {Amaro}, {Anthony}, {Arneodo}, {Barrow}, {Baudis}, {Bauermeister}, {Benabderrahmane}, {Berger}, {Breur}, {Brown}, {Brown}, {Brown}, {Bruenner}, {Bruno}, {Budnik}, {B{\"u}tikofer}, {Calv{\'e}n}, {Cardoso}, {Cervantes}, {Cichon}, {Coderre}, {Colijn}, {Conrad}, {Cussonneau}, {Decowski}, {de Perio}, {di Gangi}, {di Giovanni}, {Diglio}, {Eurin}, {Fei}, {Ferella}, {Fieguth}, {Fulgione}, {Gallo Rosso}, {Galloway}, {Gao}, {Garbini}, {Gardner}, {Geis}, {Goetzke}, {Grandi}, {Greene}, {Grignon}, {Hasterok}, {Hogenbirk}, {Howlett}, {Itay}, {Kaminsky}, {Kazama}, {Kessler}, {Kish}, {Landsman}, {Lang}, {Lellouch}, {Levinson}, {Lin}, {Lindemann}, {Lindner}, {Lombardi}, {Lopes}, {Manfredini}, {Mari{\c{s}}}, {Marrod{\'a}n Undagoitia}, {Masbou}, {Massoli}, {Masson}, {Mayani}, {Messina}, {Micheneau}, {Molinario}, {Mor{\^a}}, {Murra}, {Naganoma}, {Ni}, {Oberlack}, {Pakarha}, {Pelssers}, {Persiani}, {Piastra}, {Pienaar}, {Pizzella}, {Piro}, {Plante},
  {Priel}, {Rauch}, {Reichard}, {Reuter}, {Riedel}, {Rizzo}, {Rosendahl}, {Rupp}, {Saldanha}, {Dos Santos}, {Sartorelli}, {Scheibelhut}, {Schindler}, {Schreiner}, {Schumann}, {Scotto Lavina}, {Selvi}, {Shagin}, {Shockley}, {Silva}, {Simgen}, {Sivers}, {Stein}, {Thapa}, {Thers}, {Tiseni}, {Trinchero}, {Tunnell}, {Vargas}, {Upole}, {Wang}, {Wang}, {Wei}, {Weinheimer}, {Wulf}, {Ye}, {Zhang}, {Zhu}, \& {Xenon Collaboration}}]{2017PhRvL.119r1301A}
{Aprile}, E., {Aalbers}, J., {Agostini}, F., {et~al.} 2017, \prl, 119, 181301, \dodoi{10.1103/PhysRevLett.119.181301}

\bibitem[{{Arcadi} {et~al.}(2018){Arcadi}, {Dutra}, {Ghosh}, {Lindner}, {Mambrini}, {Pierre}, {Profumo}, \& {Queiroz}}]{2018EPJC...78..203A}
{Arcadi}, G., {Dutra}, M., {Ghosh}, P., {et~al.} 2018, European Physical Journal C, 78, 203, \dodoi{10.1140/epjc/s10052-018-5662-y}

\bibitem[{{Artale} {et~al.}(2019){Artale}, {Pedrosa}, {Tissera}, {Cataldi}, \& {Di Cintio}}]{2019A&A...622A.197A}
{Artale}, M.~C., {Pedrosa}, S.~E., {Tissera}, P.~B., {Cataldi}, P., \& {Di Cintio}, A. 2019, \aap, 622, A197, \dodoi{10.1051/0004-6361/201834096}

\bibitem[{{Benito} {et~al.}(2021){Benito}, {Iocco}, \& {Cuoco}}]{2021PDU....3200826B}
{Benito}, M., {Iocco}, F., \& {Cuoco}, A. 2021, Physics of the Dark Universe, 32, 100826, \dodoi{10.1016/j.dark.2021.100826}

\bibitem[{{Bienayme} {et~al.}(1987){Bienayme}, {Robin}, \& {Creze}}]{1987A&A...180...94B}
{Bienayme}, O., {Robin}, A.~C., \& {Creze}, M. 1987, \aap, 180, 94

\bibitem[{{Booth} {et~al.}(2009){Booth}, {de Blok}, {Jonas}, \& {Fanaroff}}]{2009arXiv0910.2935B}
{Booth}, R.~S., {de Blok}, W.~J.~G., {Jonas}, J.~L., \& {Fanaroff}, B. 2009, arXiv e-prints, arXiv:0910.2935, \dodoi{10.48550/arXiv.0910.2935}

\bibitem[{{Bosma}(2023)}]{2023arXiv230906390B}
{Bosma}, A. 2023, arXiv e-prints, arXiv:2309.06390, \dodoi{10.48550/arXiv.2309.06390}

\bibitem[{{Bouquin} {et~al.}(2018){Bouquin}, {Gil de Paz}, {Mu{\~n}oz-Mateos}, {Boissier}, {Sheth}, {Zaritsky}, {Peletier}, {Knapen}, \& {Gallego}}]{2018ApJS..234...18B}
{Bouquin}, A. Y.~K., {Gil de Paz}, A., {Mu{\~n}oz-Mateos}, J.~C., {et~al.} 2018, \apjs, 234, 18, \dodoi{10.3847/1538-4365/aaa384}

\bibitem[{{Bovy} \& {Rix}(2013)}]{2013ApJ...779..115B}
{Bovy}, J., \& {Rix}, H.-W. 2013, \apj, 779, 115, \dodoi{10.1088/0004-637X/779/2/115}

\bibitem[{{Bovy} \& {Tremaine}(2012)}]{2012ApJ...756...89B}
{Bovy}, J., \& {Tremaine}, S. 2012, \apj, 756, 89, \dodoi{10.1088/0004-637X/756/1/89}

\bibitem[{{Boylan-Kolchin} {et~al.}(2009){Boylan-Kolchin}, {Springel}, {White}, {Jenkins}, \& {Lemson}}]{2009MNRAS.398.1150B}
{Boylan-Kolchin}, M., {Springel}, V., {White}, S. D.~M., {Jenkins}, A., \& {Lemson}, G. 2009, \mnras, 398, 1150, \dodoi{10.1111/j.1365-2966.2009.15191.x}

\bibitem[{{Camps} {et~al.}(2022){Camps}, {Kapoor}, {Trcka}, {Font}, {McCarthy}, {Trayford}, \& {Baes}}]{2022MNRAS.512.2728C}
{Camps}, P., {Kapoor}, A.~U., {Trcka}, A., {et~al.} 2022, \mnras, 512, 2728, \dodoi{10.1093/mnras/stac719}

\bibitem[{{Catena} \& {Ullio}(2010)}]{2010JCAP...08..004C}
{Catena}, R., \& {Ullio}, P. 2010, \jcap, 2010, 004, \dodoi{10.1088/1475-7516/2010/08/004}

\bibitem[{{Catinella} {et~al.}(2010){Catinella}, {Schiminovich}, {Kauffmann}, {Fabello}, {Wang}, {Hummels}, {Lemonias}, {Moran}, {Wu}, {Giovanelli}, {Haynes}, {Heckman}, {Basu-Zych}, {Blanton}, {Brinchmann}, {Budav{\'a}ri}, {Gon{\c{c}}alves}, {Johnson}, {Kennicutt}, {Madore}, {Martin}, {Rich}, {Tacconi}, {Thilker}, {Wild}, \& {Wyder}}]{2010MNRAS.403..683C}
{Catinella}, B., {Schiminovich}, D., {Kauffmann}, G., {et~al.} 2010, \mnras, 403, 683, \dodoi{10.1111/j.1365-2966.2009.16180.x}

\bibitem[{{Cautun} {et~al.}(2020){Cautun}, {Ben{\'\i}tez-Llambay}, {Deason}, {Frenk}, {Fattahi}, {G{\'o}mez}, {Grand}, {Oman}, {Navarro}, \& {Simpson}}]{2020MNRAS.494.4291C}
{Cautun}, M., {Ben{\'\i}tez-Llambay}, A., {Deason}, A.~J., {et~al.} 2020, \mnras, 494, 4291, \dodoi{10.1093/mnras/staa1017}

\bibitem[{{Chakrabarti} {et~al.}(2021){Chakrabarti}, {Chang}, {Lam}, {Vigeland}, \& {Quillen}}]{Chakrabarti+21}
{Chakrabarti}, S., {Chang}, P., {Lam}, M.~T., {Vigeland}, S.~J., \& {Quillen}, A.~C. 2021, \apjl, 907, L26, \dodoi{10.3847/2041-8213/abd635}

\bibitem[{{Chakrabarti} {et~al.}(2022){Chakrabarti}, {Stevens}, {Wright}, {Rafikov}, {Chang}, {Beatty}, \& {Huber}}]{2022ApJ...928L..17C}
{Chakrabarti}, S., {Stevens}, D.~J., {Wright}, J., {et~al.} 2022, \apjl, 928, L17, \dodoi{10.3847/2041-8213/ac5c43}

\bibitem[{{Chakrabarti} {et~al.}(2020){Chakrabarti}, {Wright}, {Chang}, {Quillen}, {Craig}, {Territo}, {D'Onghia}, {Johnston}, {De Rosa}, {Huber}, {Rhode}, \& {Nielsen}}]{2020ApJ...902L..28C}
{Chakrabarti}, S., {Wright}, J., {Chang}, P., {et~al.} 2020, \apjl, 902, L28, \dodoi{10.3847/2041-8213/abb9b5}

\bibitem[{{Chowdhury} {et~al.}(2022){Chowdhury}, {Kanekar}, \& {Chengalur}}]{2022ApJ...935L...5C}
{Chowdhury}, A., {Kanekar}, N., \& {Chengalur}, J.~N. 2022, \apjl, 935, L5, \dodoi{10.3847/2041-8213/ac8150}

\bibitem[{{Chung} {et~al.}(2009){Chung}, {van Gorkom}, {Kenney}, {Crowl}, \& {Vollmer}}]{2009AJ....138.1741C}
{Chung}, A., {van Gorkom}, J.~H., {Kenney}, J. D.~P., {Crowl}, H., \& {Vollmer}, B. 2009, \aj, 138, 1741, \dodoi{10.1088/0004-6256/138/6/1741}

\bibitem[{{Corbelli} {et~al.}(2010){Corbelli}, {Lorenzoni}, {Walterbos}, {Braun}, \& {Thilker}}]{2010A&A...511A..89C}
{Corbelli}, E., {Lorenzoni}, S., {Walterbos}, R., {Braun}, R., \& {Thilker}, D. 2010, \aap, 511, A89, \dodoi{10.1051/0004-6361/200913297}

\bibitem[{{Corwin} {et~al.}(1994){Corwin}, {Buta}, \& {de Vaucouleurs}}]{1994AJ....108.2128C}
{Corwin}, Harold~G., J., {Buta}, R.~J., \& {de Vaucouleurs}, G. 1994, \aj, 108, 2128, \dodoi{10.1086/117225}

\bibitem[{{Creze} {et~al.}(1998){Creze}, {Chereul}, {Bienayme}, \& {Pichon}}]{1998A&A...329..920C}
{Creze}, M., {Chereul}, E., {Bienayme}, O., \& {Pichon}, C. 1998, \aap, 329, 920, \dodoi{10.48550/arXiv.astro-ph/9709022}

\bibitem[{{Croton} {et~al.}(2006){Croton}, {Springel}, {White}, {De Lucia}, {Frenk}, {Gao}, {Jenkins}, {Kauffmann}, {Navarro}, \& {Yoshida}}]{2006MNRAS.365...11C}
{Croton}, D.~J., {Springel}, V., {White}, S. D.~M., {et~al.} 2006, \mnras, 365, 11, \dodoi{10.1111/j.1365-2966.2005.09675.x}

\bibitem[{{Cui} {et~al.}(2021){Cui}, {Dav{\'e}}, {Peacock}, {Angl{\'e}s-Alc{\'a}zar}, \& {Yang}}]{2021NatAs...5.1069C}
{Cui}, W., {Dav{\'e}}, R., {Peacock}, J.~A., {Angl{\'e}s-Alc{\'a}zar}, D., \& {Yang}, X. 2021, Nature Astronomy, 5, 1069, \dodoi{10.1038/s41550-021-01404-1}

\bibitem[{{Dav{\'e}} {et~al.}(2019){Dav{\'e}}, {Angl{\'e}s-Alc{\'a}zar}, {Narayanan}, {Li}, {Rafieferantsoa}, \& {Appleby}}]{2019MNRAS.486.2827D}
{Dav{\'e}}, R., {Angl{\'e}s-Alc{\'a}zar}, D., {Narayanan}, D., {et~al.} 2019, \mnras, 486, 2827, \dodoi{10.1093/mnras/stz937}

\bibitem[{{Davies} {et~al.}(2017){Davies}, {Baes}, {Bianchi}, {Jones}, {Madden}, {Xilouris}, {Bocchio}, {Casasola}, {Cassara}, {Clark}, {De Looze}, {Evans}, {Fritz}, {Galametz}, {Galliano}, {Lianou}, {Mosenkov}, {Smith}, {Verstocken}, {Viaene}, {Vika}, {Wagle}, \& {Ysard}}]{DustPedia}
{Davies}, J.~I., {Baes}, M., {Bianchi}, S., {et~al.} 2017, \pasp, 129, 044102, \dodoi{10.1088/1538-3873/129/974/044102}

\bibitem[{{de Blok} {et~al.}(2008){de Blok}, {Walter}, {Brinks}, {Trachternach}, {Oh}, \& {Kennicutt}}]{2008AJ....136.2648D}
{de Blok}, W.~J.~G., {Walter}, F., {Brinks}, E., {et~al.} 2008, \aj, 136, 2648, \dodoi{10.1088/0004-6256/136/6/2648}

\bibitem[{{de Isídio} {et~al.}(2022){de Isídio}, {Men{\'e}ndez-Delmestre}, \& {Gon{\c{c}}alves}}]{2022vhow.confE..18D}
{de Isídio}, N., {Men{\'e}ndez-Delmestre}, K., \& {Gon{\c{c}}alves}, T. 2022, in VLTI-How: The VLTI High angular resolution Observations Workshop (vltihow2022, 18, \dodoi{10.5281/zenodo.7514098}

\bibitem[{{de Martino} {et~al.}(2020){de Martino}, {Chakrabarty}, {Cesare}, {Gallo}, {Ostorero}, \& {Diaferio}}]{2020Univ....6..107D}
{de Martino}, I., {Chakrabarty}, S.~S., {Cesare}, V., {et~al.} 2020, Universe, 6, 107, \dodoi{10.3390/universe6080107}

\bibitem[{{de Salas} {et~al.}(2019){de Salas}, {Malhan}, {Freese}, {Hattori}, \& {Valluri}}]{2019JCAP...10..037D}
{de Salas}, P.~F., {Malhan}, K., {Freese}, K., {Hattori}, K., \& {Valluri}, M. 2019, \jcap, 2019, 037, \dodoi{10.1088/1475-7516/2019/10/037}

\bibitem[{{Dehnen} \& {Binney}(1998)}]{1998MNRAS.294..429D}
{Dehnen}, W., \& {Binney}, J. 1998, \mnras, 294, 429, \dodoi{10.1046/j.1365-8711.1998.01282.x10.1111/j.1365-8711.1998.01282.x}

\bibitem[{{Di Teodoro}(2015)}]{2015PhDT.......216D}
{Di Teodoro}, E.~M. 2015, PhD thesis, University of Bologna, Italy

\bibitem[{{Di Teodoro} \& {Fraternali}(2015)}]{2015MNRAS.451.3021D}
{Di Teodoro}, E.~M., \& {Fraternali}, F. 2015, \mnras, 451, 3021, \dodoi{10.1093/mnras/stv1213}

\bibitem[{{Di Teodoro} {et~al.}(2023){Di Teodoro}, {Posti}, {Fall}, {Ogle}, {Jarrett}, {Appleton}, {Cluver}, {Haynes}, \& {Lisenfeld}}]{2023MNRAS.518.6340D}
{Di Teodoro}, E.~M., {Posti}, L., {Fall}, S.~M., {et~al.} 2023, \mnras, 518, 6340, \dodoi{10.1093/mnras/stac3424}

\bibitem[{{Dutton} \& {Macci{\`o}}(2014)}]{2014MNRAS.441.3359D}
{Dutton}, A.~A., \& {Macci{\`o}}, A.~V. 2014, \mnras, 441, 3359, \dodoi{10.1093/mnras/stu742}

\bibitem[{{Eilers} {et~al.}(2019){Eilers}, {Hogg}, {Rix}, \& {Ness}}]{2019ApJ...871..120E}
{Eilers}, A.-C., {Hogg}, D.~W., {Rix}, H.-W., \& {Ness}, M.~K. 2019, \apj, 871, 120, \dodoi{10.3847/1538-4357/aaf648}

\bibitem[{{Einasto}(1965)}]{Einasto}
{Einasto}, J. 1965, Trudy Astrofizicheskogo Instituta Alma-Ata, 5, 87

\bibitem[{{Enia} {et~al.}(2020){Enia}, {Rodighiero}, {Morselli}, {Casasola}, {Bianchi}, {Rodriguez-Mu{\~n}oz}, {Mancini}, {Renzini}, {Popesso}, {Cassata}, {Negrello}, \& {Franceschini}}]{SED_Fitting}
{Enia}, A., {Rodighiero}, G., {Morselli}, L., {et~al.} 2020, \mnras, 493, 4107, \dodoi{10.1093/mnras/staa433}

\bibitem[{{Eskew} {et~al.}(2012){Eskew}, {Zaritsky}, \& {Meidt}}]{2012AJ....143..139E}
{Eskew}, M., {Zaritsky}, D., \& {Meidt}, S. 2012, \aj, 143, 139, \dodoi{10.1088/0004-6256/143/6/139}

\bibitem[{{Fazio} {et~al.}(2004){Fazio}, {Hora}, {Allen}, {Ashby}, {Barmby}, {Deutsch}, {Huang}, {Kleiner}, {Marengo}, {Megeath}, {Melnick}, {Pahre}, {Patten}, {Polizotti}, {Smith}, {Taylor}, {Wang}, {Willner}, {Hoffmann}, {Pipher}, {Forrest}, {McMurty}, {McCreight}, {McKelvey}, {McMurray}, {Koch}, {Moseley}, {Arendt}, {Mentzell}, {Marx}, {Losch}, {Mayman}, {Eichhorn}, {Krebs}, {Jhabvala}, {Gezari}, {Fixsen}, {Flores}, {Shakoorzadeh}, {Jungo}, {Hakun}, {Workman}, {Karpati}, {Kichak}, {Whitley}, {Mann}, {Tollestrup}, {Eisenhardt}, {Stern}, {Gorjian}, {Bhattacharya}, {Carey}, {Nelson}, {Glaccum}, {Lacy}, {Lowrance}, {Laine}, {Reach}, {Stauffer}, {Surace}, {Wilson}, {Wright}, {Hoffman}, {Domingo}, \& {Cohen}}]{2004ApJS..154...10F}
{Fazio}, G.~G., {Hora}, J.~L., {Allen}, L.~E., {et~al.} 2004, \apjs, 154, 10, \dodoi{10.1086/422843}

\bibitem[{{Foreman-Mackey} {et~al.}(2013){Foreman-Mackey}, {Hogg}, {Lang}, \& {Goodman}}]{2013PASP..125..306F}
{Foreman-Mackey}, D., {Hogg}, D.~W., {Lang}, D., \& {Goodman}, J. 2013, \pasp, 125, 306, \dodoi{10.1086/670067}

\bibitem[{{Frank} {et~al.}(2016){Frank}, {de Blok}, {Walter}, {Leroy}, \& {Carignan}}]{2016AJ....151...94F}
{Frank}, B.~S., {de Blok}, W.~J.~G., {Walter}, F., {Leroy}, A., \& {Carignan}, C. 2016, \aj, 151, 94, \dodoi{10.3847/0004-6256/151/4/94}

\bibitem[{{Georgantopoulos} \& {Zezas}(2003)}]{2003ApJ...594..704G}
{Georgantopoulos}, I., \& {Zezas}, A. 2003, \apj, 594, 704, \dodoi{10.1086/377120}

\bibitem[{{Girelli} {et~al.}(2020){Girelli}, {Pozzetti}, {Bolzonella}, {Giocoli}, {Marulli}, \& {Baldi}}]{2020A&A...634A.135G}
{Girelli}, G., {Pozzetti}, L., {Bolzonella}, M., {et~al.} 2020, \aap, 634, A135, \dodoi{10.1051/0004-6361/201936329}

\bibitem[{{Goodwin} {et~al.}(1998){Goodwin}, {Gribbin}, \& {Hendry}}]{1998Obs...118..201G}
{Goodwin}, S.~P., {Gribbin}, J., \& {Hendry}, M.~A. 1998, The Observatory, 118, 201

\bibitem[{{Holmberg} \& {Flynn}(2004)}]{2004MNRAS.352..440H}
{Holmberg}, J., \& {Flynn}, C. 2004, \mnras, 352, 440, \dodoi{10.1111/j.1365-2966.2004.07931.x}

\bibitem[{{Hopkins} {et~al.}(2012){Hopkins}, {Quataert}, \& {Murray}}]{2012MNRAS.421.3522H}
{Hopkins}, P.~F., {Quataert}, E., \& {Murray}, N. 2012, \mnras, 421, 3522, \dodoi{10.1111/j.1365-2966.2012.20593.x}

\bibitem[{{Hou} \& {Han}(2014)}]{2014A&A...569A.125H}
{Hou}, L.~G., \& {Han}, J.~L. 2014, \aap, 569, A125, \dodoi{10.1051/0004-6361/201424039}

\bibitem[{{Huang} {et~al.}(2016){Huang}, {Liu}, {Yuan}, {Xiang}, {Zhang}, {Chen}, {Ren}, {Wang}, {Zhang}, {Hou}, {Wang}, \& {Cao}}]{2016MNRAS.463.2623H}
{Huang}, Y., {Liu}, X.~W., {Yuan}, H.~B., {et~al.} 2016, \mnras, 463, 2623, \dodoi{10.1093/mnras/stw2096}

\bibitem[{{Iocco} {et~al.}(2011){Iocco}, {Pato}, {Bertone}, \& {Jetzer}}]{2011JCAP...11..029I}
{Iocco}, F., {Pato}, M., {Bertone}, G., \& {Jetzer}, P. 2011, \jcap, 2011, 029, \dodoi{10.1088/1475-7516/2011/11/029}

\bibitem[{{Jarrett} {et~al.}(2023){Jarrett}, {Cluver}, {Taylor}, {Bellstedt}, {Robotham}, \& {Yao}}]{2023arXiv230105952J}
{Jarrett}, T.~H., {Cluver}, M.~E., {Taylor}, E.~N., {et~al.} 2023, arXiv e-prints, arXiv:2301.05952, \dodoi{10.48550/arXiv.2301.05952}

\bibitem[{{Jeans}(1922)}]{1922MNRAS..82..122J}
{Jeans}, J.~H. 1922, \mnras, 82, 122, \dodoi{10.1093/mnras/82.3.122}

\bibitem[{{Jones} {et~al.}(2021){Jones}, {Vergani}, {Romano}, {Ginolfi}, {Fudamoto}, {B{\'e}thermin}, {Fujimoto}, {Lemaux}, {Morselli}, {Capak}, {Cassata}, {Faisst}, {Le F{\`e}vre}, {Schaerer}, {Silverman}, {Yan}, {Boquien}, {Cimatti}, {Dessauges-Zavadsky}, {Ibar}, {Maiolino}, {Rizzo}, {Talia}, \& {Zamorani}}]{2021MNRAS.507.3540J}
{Jones}, G.~C., {Vergani}, D., {Romano}, M., {et~al.} 2021, \mnras, 507, 3540, \dodoi{10.1093/mnras/stab2226}

\bibitem[{{Kapteyn}(1922)}]{1922ApJ....55..302K}
{Kapteyn}, J.~C. 1922, \apj, 55, 302, \dodoi{10.1086/142670}

\bibitem[{{Karukes} {et~al.}(2019){Karukes}, {Benito}, {Iocco}, {Trotta}, \& {Geringer-Sameth}}]{2019JCAP...09..046K}
{Karukes}, E.~V., {Benito}, M., {Iocco}, F., {Trotta}, R., \& {Geringer-Sameth}, A. 2019, \jcap, 2019, 046, \dodoi{10.1088/1475-7516/2019/09/046}

\bibitem[{{Kormendy} \& {Freeman}(2016)}]{2016ApJ...817...84K}
{Kormendy}, J., \& {Freeman}, K.~C. 2016, \apj, 817, 84, \dodoi{10.3847/0004-637X/817/2/84}

\bibitem[{{Kravtsov} {et~al.}(2018){Kravtsov}, {Vikhlinin}, \& {Meshcheryakov}}]{2018AstL...44....8K}
{Kravtsov}, A.~V., {Vikhlinin}, A.~A., \& {Meshcheryakov}, A.~V. 2018, Astronomy Letters, 44, 8, \dodoi{10.1134/S1063773717120015}

\bibitem[{{Lazio}(2009)}]{2009pra..confE..58L}
{Lazio}, J. 2009, in Panoramic Radio Astronomy: Wide-field 1-2 GHz Research on Galaxy Evolution, 58, \dodoi{10.22323/1.089.0058}

\bibitem[{{Lelli} {et~al.}(2016){Lelli}, {McGaugh}, \& {Schombert}}]{2016AJ....152..157L}
{Lelli}, F., {McGaugh}, S.~S., \& {Schombert}, J.~M. 2016, \aj, 152, 157, \dodoi{10.3847/0004-6256/152/6/157}

\bibitem[{{Lewin} \& {Smith}(1996)}]{1996APh.....6...87L}
{Lewin}, J.~D., \& {Smith}, P.~F. 1996, Astroparticle Physics, 6, 87, \dodoi{10.1016/S0927-6505(96)00047-3}

\bibitem[{{Li} {et~al.}(2020){Li}, {Lelli}, {McGaugh}, \& {Schombert}}]{2020ApJS..247...31L}
{Li}, P., {Lelli}, F., {McGaugh}, S., \& {Schombert}, J. 2020, \apjs, 247, 31, \dodoi{10.3847/1538-4365/ab700e}

\bibitem[{{Licquia} \& {Newman}(2015)}]{2015ApJ...806...96L}
{Licquia}, T.~C., \& {Newman}, J.~A. 2015, \apj, 806, 96, \dodoi{10.1088/0004-637X/806/1/96}

\bibitem[{{Lin} \& {Li}(2019)}]{2019MNRAS.487.5679L}
{Lin}, H.-N., \& {Li}, X. 2019, \mnras, 487, 5679, \dodoi{10.1093/mnras/stz1698}

\bibitem[{{Makarov} {et~al.}(2014){Makarov}, {Prugniel}, {Terekhova}, {Courtois}, \& {Vauglin}}]{2014A&A...570A..13M}
{Makarov}, D., {Prugniel}, P., {Terekhova}, N., {Courtois}, H., \& {Vauglin}, I. 2014, \aap, 570, A13, \dodoi{10.1051/0004-6361/201423496}

\bibitem[{{Mancera Pi{\~n}a} {et~al.}(2022){Mancera Pi{\~n}a}, {Fraternali}, {Oosterloo}, {Adams}, {di Teodoro}, {Bacchini}, \& {Iorio}}]{2022MNRAS.514.3329M}
{Mancera Pi{\~n}a}, P.~E., {Fraternali}, F., {Oosterloo}, T., {et~al.} 2022, \mnras, 514, 3329, \dodoi{10.1093/mnras/stac1508}

\bibitem[{{McMillan}(2011)}]{2011MNRAS.414.2446M}
{McMillan}, P.~J. 2011, \mnras, 414, 2446, \dodoi{10.1111/j.1365-2966.2011.18564.x}

\bibitem[{{McMillan}(2017)}]{2017MNRAS.465...76M}
---. 2017, \mnras, 465, 76, \dodoi{10.1093/mnras/stw2759}

\bibitem[{{Merrifield}(1992)}]{1992AJ....103.1552M}
{Merrifield}, M.~R. 1992, \aj, 103, 1552, \dodoi{10.1086/116168}

\bibitem[{{Mo} {et~al.}(2010){Mo}, {van den Bosch}, \& {White}}]{2010gfe..book.....M}
{Mo}, H., {van den Bosch}, F.~C., \& {White}, S. 2010, {Galaxy Formation and Evolution}

\bibitem[{{More} {et~al.}(2011){More}, {van den Bosch}, {Cacciato}, {Skibba}, {Mo}, \& {Yang}}]{2011MNRAS.410..210M}
{More}, S., {van den Bosch}, F.~C., {Cacciato}, M., {et~al.} 2011, \mnras, 410, 210, \dodoi{10.1111/j.1365-2966.2010.17436.x}

\bibitem[{{Moster} {et~al.}(2013){Moster}, {Naab}, \& {White}}]{2013MNRAS.428.3121M}
{Moster}, B.~P., {Naab}, T., \& {White}, S. D.~M. 2013, \mnras, 428, 3121, \dodoi{10.1093/mnras/sts261}

\bibitem[{{Mu{\~n}oz-Mateos} {et~al.}(2013){Mu{\~n}oz-Mateos}, {Sheth}, {Gil de Paz}, {Meidt}, {Athanassoula}, {Bosma}, {Comer{\'o}n}, {Elmegreen}, {Elmegreen}, {Erroz-Ferrer}, {Gadotti}, {Hinz}, {Ho}, {Holwerda}, {Jarrett}, {Kim}, {Knapen}, {Laine}, {Laurikainen}, {Madore}, {Menendez-Delmestre}, {Mizusawa}, {Regan}, {Salo}, {Schinnerer}, {Seibert}, {Skibba}, \& {Zaritsky}}]{2013ApJ...771...59M}
{Mu{\~n}oz-Mateos}, J.~C., {Sheth}, K., {Gil de Paz}, A., {et~al.} 2013, \apj, 771, 59, \dodoi{10.1088/0004-637X/771/1/59}

\bibitem[{{Mu{\~n}oz-Mateos} {et~al.}(2015){Mu{\~n}oz-Mateos}, {Sheth}, {Regan}, {Kim}, {Laine}, {Erroz-Ferrer}, {Gil de Paz}, {Comeron}, {Hinz}, {Laurikainen}, {Salo}, {Athanassoula}, {Bosma}, {Bouquin}, {Schinnerer}, {Ho}, {Zaritsky}, {Gadotti}, {Madore}, {Holwerda}, {Men{\'e}ndez-Delmestre}, {Knapen}, {Meidt}, {Querejeta}, {Mizusawa}, {Seibert}, {Laine}, \& {Courtois}}]{2015ApJS..219....3M}
{Mu{\~n}oz-Mateos}, J.~C., {Sheth}, K., {Regan}, M., {et~al.} 2015, \apjs, 219, 3, \dodoi{10.1088/0067-0049/219/1/3}

\bibitem[{{Navarro} {et~al.}(1996){Navarro}, {Frenk}, \& {White}}]{1996ApJ...462..563N}
{Navarro}, J.~F., {Frenk}, C.~S., \& {White}, S. D.~M. 1996, \apj, 462, 563, \dodoi{10.1086/177173}

\bibitem[{{Navarro} {et~al.}(1997){Navarro}, {Frenk}, \& {White}}]{1997ApJ...490..493N}
---. 1997, \apj, 490, 493, \dodoi{10.1086/304888}

\bibitem[{{Navarro} {et~al.}(2010){Navarro}, {Ludlow}, {Springel}, {Wang}, {Vogelsberger}, {White}, {Jenkins}, {Frenk}, \& {Helmi}}]{2010MNRAS.402...21N}
{Navarro}, J.~F., {Ludlow}, A., {Springel}, V., {et~al.} 2010, \mnras, 402, 21, \dodoi{10.1111/j.1365-2966.2009.15878.x}

\bibitem[{{Oort}(1932)}]{1932BAN.....6..249O}
{Oort}, J.~H. 1932, \bain, 6, 249

\bibitem[{{Palau} \& {Miralda-Escud{\'e}}(2023)}]{2023MNRAS.524.2124P}
{Palau}, C.~G., \& {Miralda-Escud{\'e}}, J. 2023, \mnras, 524, 2124, \dodoi{10.1093/mnras/stad1930}

\bibitem[{{Pappalardo} {et~al.}(2012){Pappalardo}, {Bianchi}, {Corbelli}, {Giovanardi}, {Hunt}, {Bendo}, {Boselli}, {Cortese}, {Magrini}, {Zibetti}, {di Serego Alighieri}, {Davies}, {Baes}, {Ciesla}, {Clemens}, {De Looze}, {Fritz}, {Grossi}, {Pohlen}, {Smith}, {Verstappen}, \& {Vlahakis}}]{2012A&A...545A..75P}
{Pappalardo}, C., {Bianchi}, S., {Corbelli}, E., {et~al.} 2012, \aap, 545, A75, \dodoi{10.1051/0004-6361/201219689}

\bibitem[{{Pato} {et~al.}(2015){Pato}, {Iocco}, \& {Bertone}}]{2015JCAP...12..001P}
{Pato}, M., {Iocco}, F., \& {Bertone}, G. 2015, \jcap, 2015, 001, \dodoi{10.1088/1475-7516/2015/12/001}

\bibitem[{{Peters} {et~al.}(2017){Peters}, {van der Kruit}, {Allen}, \& {Freeman}}]{2017MNRAS.464...65P}
{Peters}, S.~P.~C., {van der Kruit}, P.~C., {Allen}, R.~J., \& {Freeman}, K.~C. 2017, \mnras, 464, 65, \dodoi{10.1093/mnras/stw2101}

\bibitem[{{Piffl} {et~al.}(2014){Piffl}, {Binney}, {McMillan}, {Steinmetz}, {Helmi}, {Wyse}, {Bienaym{\'e}}, {Bland-Hawthorn}, {Freeman}, {Gibson}, {Gilmore}, {Grebel}, {Kordopatis}, {Navarro}, {Parker}, {Reid}, {Seabroke}, {Siebert}, {Watson}, \& {Zwitter}}]{2014MNRAS.445.3133P}
{Piffl}, T., {Binney}, J., {McMillan}, P.~J., {et~al.} 2014, \mnras, 445, 3133, \dodoi{10.1093/mnras/stu1948}

\bibitem[{{Pillepich} {et~al.}(2023){Pillepich}, {Sotillo-Ramos}, {Ramesh}, {Nelson}, {Engler}, {Rodriguez-Gomez}, {Fournier}, {Donnari}, {Springel}, \& {Hernquist}}]{2023arXiv230316217P}
{Pillepich}, A., {Sotillo-Ramos}, D., {Ramesh}, R., {et~al.} 2023, arXiv e-prints, arXiv:2303.16217, \dodoi{10.48550/arXiv.2303.16217}

\bibitem[{{Planck Collaboration} {et~al.}(2016){Planck Collaboration}, {Ade}, {Aghanim}, {Arnaud}, {Ashdown}, {Aumont}, {Baccigalupi}, {Banday}, {Barreiro}, {Bartlett}, {Bartolo}, {Battaner}, {Battye}, {Benabed}, {Beno{\^\i}t}, {Benoit-L{\'e}vy}, {Bernard}, {Bersanelli}, {Bielewicz}, {Bock}, {Bonaldi}, {Bonavera}, {Bond}, {Borrill}, {Bouchet}, {Boulanger}, {Bucher}, {Burigana}, {Butler}, {Calabrese}, {Cardoso}, {Catalano}, {Challinor}, {Chamballu}, {Chary}, {Chiang}, {Chluba}, {Christensen}, {Church}, {Clements}, {Colombi}, {Colombo}, {Combet}, {Coulais}, {Crill}, {Curto}, {Cuttaia}, {Danese}, {Davies}, {Davis}, {de Bernardis}, {de Rosa}, {de Zotti}, {Delabrouille}, {D{\'e}sert}, {Di Valentino}, {Dickinson}, {Diego}, {Dolag}, {Dole}, {Donzelli}, {Dor{\'e}}, {Douspis}, {Ducout}, {Dunkley}, {Dupac}, {Efstathiou}, {Elsner}, {En{\ss}lin}, {Eriksen}, {Farhang}, {Fergusson}, {Finelli}, {Forni}, {Frailis}, {Fraisse}, {Franceschi}, {Frejsel}, {Galeotta}, {Galli}, {Ganga}, {Gauthier}, {Gerbino}, {Ghosh}, {Giard},
  {Giraud-H{\'e}raud}, {Giusarma}, {Gjerl{\o}w}, {Gonz{\'a}lez-Nuevo}, {G{\'o}rski}, {Gratton}, {Gregorio}, {Gruppuso}, {Gudmundsson}, {Hamann}, {Hansen}, {Hanson}, {Harrison}, {Helou}, {Henrot-Versill{\'e}}, {Hern{\'a}ndez-Monteagudo}, {Herranz}, {Hildebrandt}, {Hivon}, {Hobson}, {Holmes}, {Hornstrup}, {Hovest}, {Huang}, {Huffenberger}, {Hurier}, {Jaffe}, {Jaffe}, {Jones}, {Juvela}, {Keih{\"a}nen}, {Keskitalo}, {Kisner}, {Kneissl}, {Knoche}, {Knox}, {Kunz}, {Kurki-Suonio}, {Lagache}, {L{\"a}hteenm{\"a}ki}, {Lamarre}, {Lasenby}, {Lattanzi}, {Lawrence}, {Leahy}, {Leonardi}, {Lesgourgues}, {Levrier}, {Lewis}, {Liguori}, {Lilje}, {Linden-V{\o}rnle}, {L{\'o}pez-Caniego}, {Lubin}, {Mac{\'\i}as-P{\'e}rez}, {Maggio}, {Maino}, {Mandolesi}, {Mangilli}, {Marchini}, {Maris}, {Martin}, {Martinelli}, {Mart{\'\i}nez-Gonz{\'a}lez}, {Masi}, {Matarrese}, {McGehee}, {Meinhold}, {Melchiorri}, {Melin}, {Mendes}, {Mennella}, {Migliaccio}, {Millea}, {Mitra}, {Miville-Desch{\^e}nes}, {Moneti}, {Montier}, {Morgante}, {Mortlock},
  {Moss}, {Munshi}, {Murphy}, {Naselsky}, {Nati}, {Natoli}, {Netterfield}, {N{\o}rgaard-Nielsen}, {Noviello}, {Novikov}, {Novikov}, {Oxborrow}, {Paci}, {Pagano}, {Pajot}, {Paladini}, {Paoletti}, {Partridge}, {Pasian}, {Patanchon}, {Pearson}, {Perdereau}, {Perotto}, {Perrotta}, {Pettorino}, {Piacentini}, {Piat}, {Pierpaoli}, {Pietrobon}, {Plaszczynski}, {Pointecouteau}, {Polenta}, {Popa}, {Pratt}, {Pr{\'e}zeau}, {Prunet}, {Puget}, {Rachen}, {Reach}, {Rebolo}, {Reinecke}, {Remazeilles}, {Renault}, {Renzi}, {Ristorcelli}, {Rocha}, {Rosset}, {Rossetti}, {Roudier}, {Rouill{\'e} d'Orfeuil}, {Rowan-Robinson}, {Rubi{\~n}o-Mart{\'\i}n}, {Rusholme}, {Said}, {Salvatelli}, {Salvati}, {Sandri}, {Santos}, {Savelainen}, {Savini}, {Scott}, {Seiffert}, {Serra}, {Shellard}, {Spencer}, {Spinelli}, {Stolyarov}, {Stompor}, {Sudiwala}, {Sunyaev}, {Sutton}, {Suur-Uski}, {Sygnet}, {Tauber}, {Terenzi}, {Toffolatti}, {Tomasi}, {Tristram}, {Trombetti}, {Tucci}, {Tuovinen}, {T{\"u}rler}, {Umana}, {Valenziano}, {Valiviita}, {Van Tent},
  {Vielva}, {Villa}, {Wade}, {Wandelt}, {Wehus}, {White}, {White}, {Wilkinson}, {Yvon}, {Zacchei}, \& {Zonca}}]{2016A&A...594A..13P}
{Planck Collaboration}, {Ade}, P.~A.~R., {Aghanim}, N., {et~al.} 2016, \aap, 594, A13, \dodoi{10.1051/0004-6361/201525830}

\bibitem[{{Planck Collaboration} {et~al.}(2020){Planck Collaboration}, {Akrami}, {Arroja}, {Ashdown}, {Aumont}, {Baccigalupi}, {Ballardini}, {Banday}, {Barreiro}, {Bartolo}, {Basak}, {Benabed}, {Bernard}, {Bersanelli}, {Bielewicz}, {Bock}, {Bond}, {Borrill}, {Bouchet}, {Boulanger}, {Bucher}, {Burigana}, {Butler}, {Calabrese}, {Cardoso}, {Carron}, {Challinor}, {Chiang}, {Colombo}, {Combet}, {Contreras}, {Crill}, {Cuttaia}, {de Bernardis}, {de Zotti}, {Delabrouille}, {Delouis}, {Di Valentino}, {Diego}, {Donzelli}, {Dor{\'e}}, {Douspis}, {Ducout}, {Dupac}, {Dusini}, {Efstathiou}, {Elsner}, {En{\ss}lin}, {Eriksen}, {Fantaye}, {Fergusson}, {Fernandez-Cobos}, {Finelli}, {Forastieri}, {Frailis}, {Franceschi}, {Frolov}, {Galeotta}, {Galli}, {Ganga}, {Gauthier}, {G{\'e}nova-Santos}, {Gerbino}, {Ghosh}, {Gonz{\'a}lez-Nuevo}, {G{\'o}rski}, {Gratton}, {Gruppuso}, {Gudmundsson}, {Hamann}, {Handley}, {Hansen}, {Herranz}, {Hivon}, {Hooper}, {Huang}, {Jaffe}, {Jones}, {Keih{\"a}nen}, {Keskitalo}, {Kiiveri}, {Kim}, {Kisner},
  {Krachmalnicoff}, {Kunz}, {Kurki-Suonio}, {Lagache}, {Lamarre}, {Lasenby}, {Lattanzi}, {Lawrence}, {Le Jeune}, {Lesgourgues}, {Levrier}, {Lewis}, {Liguori}, {Lilje}, {Lindholm}, {L{\'o}pez-Caniego}, {Lubin}, {Ma}, {Mac{\'\i}as-P{\'e}rez}, {Maggio}, {Maino}, {Mandolesi}, {Mangilli}, {Marcos-Caballero}, {Maris}, {Martin}, {Mart{\'\i}nez-Gonz{\'a}lez}, {Matarrese}, {Mauri}, {McEwen}, {Meerburg}, {Meinhold}, {Melchiorri}, {Mennella}, {Migliaccio}, {Mitra}, {Miville-Desch{\^e}nes}, {Molinari}, {Moneti}, {Montier}, {Morgante}, {Moss}, {M{\"u}nchmeyer}, {Natoli}, {N{\o}rgaard-Nielsen}, {Pagano}, {Paoletti}, {Partridge}, {Patanchon}, {Peiris}, {Perrotta}, {Pettorino}, {Piacentini}, {Polastri}, {Polenta}, {Puget}, {Rachen}, {Reinecke}, {Remazeilles}, {Renzi}, {Rocha}, {Rosset}, {Roudier}, {Rubi{\~n}o-Mart{\'\i}n}, {Ruiz-Granados}, {Salvati}, {Sandri}, {Savelainen}, {Scott}, {Shellard}, {Shiraishi}, {Sirignano}, {Sirri}, {Spencer}, {Sunyaev}, {Suur-Uski}, {Tauber}, {Tavagnacco}, {Tenti}, {Toffolatti}, {Tomasi},
  {Trombetti}, {Valiviita}, {Van Tent}, {Vielva}, {Villa}, {Vittorio}, {Wandelt}, {Wehus}, {White}, {Zacchei}, {Zibin}, \& {Zonca}}]{2020A&A...641A..10P}
{Planck Collaboration}, {Akrami}, Y., {Arroja}, F., {et~al.} 2020, \aap, 641, A10, \dodoi{10.1051/0004-6361/201833887}

\bibitem[{{Pogge} {et~al.}(2000){Pogge}, {Maoz}, {Ho}, \& {Eracleous}}]{2000ApJ...532..323P}
{Pogge}, R.~W., {Maoz}, D., {Ho}, L.~C., \& {Eracleous}, M. 2000, \apj, 532, 323, \dodoi{10.1086/308567}

\bibitem[{{Posti} \& {Fall}(2021)}]{2021A&A...649A.119P}
{Posti}, L., \& {Fall}, S.~M. 2021, \aap, 649, A119, \dodoi{10.1051/0004-6361/202040256}

\bibitem[{{Posti} {et~al.}(2019){Posti}, {Fraternali}, \& {Marasco}}]{2019A&A...626A..56P}
{Posti}, L., {Fraternali}, F., \& {Marasco}, A. 2019, \aap, 626, A56, \dodoi{10.1051/0004-6361/201935553}

\bibitem[{{Price} {et~al.}(2021){Price}, {Shimizu}, {Genzel}, {{\"U}bler}, {F{\"o}rster Schreiber}, {Tacconi}, {Davies}, {Coogan}, {Lutz}, {Wuyts}, {Wisnioski}, {Nestor}, {Sternberg}, {Burkert}, {Bender}, {Contursi}, {Davies}, {Herrera-Camus}, {Lee}, {Naab}, {Neri}, {Renzini}, {Saglia}, {Schruba}, \& {Schuster}}]{2021ApJ...922..143P}
{Price}, S.~H., {Shimizu}, T.~T., {Genzel}, R., {et~al.} 2021, \apj, 922, 143, \dodoi{10.3847/1538-4357/ac22ad}

\bibitem[{{Querejeta} {et~al.}(2015){Querejeta}, {Meidt}, {Schinnerer}, {Cisternas}, {Mu{\~n}oz-Mateos}, {Sheth}, {Knapen}, {van de Ven}, {Norris}, {Peletier}, {Laurikainen}, {Salo}, {Holwerda}, {Athanassoula}, {Bosma}, {Groves}, {Ho}, {Gadotti}, {Zaritsky}, {Regan}, {Hinz}, {Gil de Paz}, {Menendez-Delmestre}, {Seibert}, {Mizusawa}, {Kim}, {Erroz-Ferrer}, {Laine}, \& {Comer{\'o}n}}]{2015ApJS..219....5Q}
{Querejeta}, M., {Meidt}, S.~E., {Schinnerer}, E., {et~al.} 2015, \apjs, 219, 5, \dodoi{10.1088/0067-0049/219/1/5}

\bibitem[{{Quiroga-Nu{\~n}ez} {et~al.}(2017){Quiroga-Nu{\~n}ez}, {van Langevelde}, {Reid}, \& {Green}}]{2017A&A...604A..72Q}
{Quiroga-Nu{\~n}ez}, L.~H., {van Langevelde}, H.~J., {Reid}, M.~J., \& {Green}, J.~A. 2017, \aap, 604, A72, \dodoi{10.1051/0004-6361/201730681}

\bibitem[{{Randriamampandry} {et~al.}(2021){Randriamampandry}, {Wang}, \& {Mogotsi}}]{2021ApJ...916...26R}
{Randriamampandry}, T.~H., {Wang}, J., \& {Mogotsi}, K.~M. 2021, \apj, 916, 26, \dodoi{10.3847/1538-4357/ac0442}

\bibitem[{{Read}(2014)}]{2014JPhG...41f3101R}
{Read}, J.~I. 2014, Journal of Physics G Nuclear Physics, 41, 063101, \dodoi{10.1088/0954-3899/41/6/063101}

\bibitem[{{Reid} \& {Dame}(2016)}]{2016ApJ...832..159R}
{Reid}, M.~J., \& {Dame}, T.~M. 2016, \apj, 832, 159, \dodoi{10.3847/0004-637X/832/2/159}

\bibitem[{{Reid} {et~al.}(2019){Reid}, {Menten}, {Brunthaler}, {Zheng}, {Dame}, {Xu}, {Li}, {Sakai}, {Wu}, {Immer}, {Zhang}, {Sanna}, {Moscadelli}, {Rygl}, {Bartkiewicz}, {Hu}, {Quiroga-Nu{\~n}ez}, \& {van Langevelde}}]{2019ApJ...885..131R}
{Reid}, M.~J., {Menten}, K.~M., {Brunthaler}, A., {et~al.} 2019, \apj, 885, 131, \dodoi{10.3847/1538-4357/ab4a11}

\bibitem[{{Roman-Oliveira} {et~al.}(2023){Roman-Oliveira}, {Fraternali}, \& {Rizzo}}]{2023MNRAS.521.1045R}
{Roman-Oliveira}, F., {Fraternali}, F., \& {Rizzo}, F. 2023, \mnras, 521, 1045, \dodoi{10.1093/mnras/stad530}

\bibitem[{{Romeo} {et~al.}(2020){Romeo}, {Agertz}, \& {Renaud}}]{2020MNRAS.499.5656R}
{Romeo}, A.~B., {Agertz}, O., \& {Renaud}, F. 2020, \mnras, 499, 5656, \dodoi{10.1093/mnras/staa3245}

\bibitem[{{Rubin} \& {Ford}(1970)}]{1970ApJ...159..379R}
{Rubin}, V.~C., \& {Ford}, W.~Kent, J. 1970, \apj, 159, 379, \dodoi{10.1086/150317}

\bibitem[{{S4G Team}(2020)}]{https://doi.org/10.26131/irsa425}
{S4G Team}. 2020, Spitzer Survey of Stellar Structure in Galaxies,  IPAC, \dodoi{10.26131/IRSA425}

\bibitem[{{Salucci}(2019)}]{2019A&ARv..27....2S}
{Salucci}, P. 2019, \aapr, 27, 2, \dodoi{10.1007/s00159-018-0113-1}

\bibitem[{{Savastano} {et~al.}(2019){Savastano}, {Amendola}, {Rubio}, \& {Wetterich}}]{2019PhRvD.100h3518S}
{Savastano}, S., {Amendola}, L., {Rubio}, J., \& {Wetterich}, C. 2019, \prd, 100, 083518, \dodoi{10.1103/PhysRevD.100.083518}

\bibitem[{{Sharma} {et~al.}(2021){Sharma}, {Salucci}, {Harrison}, {van de Ven}, \& {Lapi}}]{2021MNRAS.503.1753S}
{Sharma}, G., {Salucci}, P., {Harrison}, C.~M., {van de Ven}, G., \& {Lapi}, A. 2021, \mnras, 503, 1753, \dodoi{10.1093/mnras/stab249}

\bibitem[{{Sheth} {et~al.}(2010){Sheth}, {Regan}, {Hinz}, {Gil de Paz}, {Men{\'e}ndez-Delmestre}, {Mu{\~n}oz-Mateos}, {Seibert}, {Kim}, {Laurikainen}, {Salo}, {Gadotti}, {Laine}, {Mizusawa}, {Armus}, {Athanassoula}, {Bosma}, {Buta}, {Capak}, {Jarrett}, {Elmegreen}, {Elmegreen}, {Knapen}, {Koda}, {Helou}, {Ho}, {Madore}, {Masters}, {Mobasher}, {Ogle}, {Peng}, {Schinnerer}, {Surace}, {Zaritsky}, {Comer{\'o}n}, {de Swardt}, {Meidt}, {Kasliwal}, \& {Aravena}}]{2010PASP..122.1397S}
{Sheth}, K., {Regan}, M., {Hinz}, J.~L., {et~al.} 2010, \pasp, 122, 1397, \dodoi{10.1086/657638}

\bibitem[{{Shin} {et~al.}(2022){Shin}, {Lee}, {Hwang}, {Song}, {Ko}, {Smith}, {Kim}, \& {Yoo}}]{2022ApJ...934...43S}
{Shin}, J., {Lee}, J.~C., {Hwang}, H.~S., {et~al.} 2022, \apj, 934, 43, \dodoi{10.3847/1538-4357/ac7961}

\bibitem[{{Shuntov} {et~al.}(2022){Shuntov}, {McCracken}, {Gavazzi}, {Laigle}, {Weaver}, {Davidzon}, {Ilbert}, {Kauffmann}, {Faisst}, {Dubois}, {Koekemoer}, {Moneti}, {Milvang-Jensen}, {Mobasher}, {Sanders}, \& {Toft}}]{2022A&A...664A..61S}
{Shuntov}, M., {McCracken}, H.~J., {Gavazzi}, R., {et~al.} 2022, \aap, 664, A61, \dodoi{10.1051/0004-6361/202243136}

\bibitem[{{Silk} \& {Rees}(1998)}]{1998A&A...331L...1S}
{Silk}, J., \& {Rees}, M.~J. 1998, \aap, 331, L1, \dodoi{10.48550/arXiv.astro-ph/9801013}

\bibitem[{{Silverwood} \& {Easther}(2019)}]{2019PASA...36...38S}
{Silverwood}, H., \& {Easther}, R. 2019, \pasa, 36, e038, \dodoi{10.1017/pasa.2019.25}

\bibitem[{{Simon} {et~al.}(2003){Simon}, {Bolatto}, {Leroy}, \& {Blitz}}]{2003ApJ...596..957S}
{Simon}, J.~D., {Bolatto}, A.~D., {Leroy}, A., \& {Blitz}, L. 2003, \apj, 596, 957, \dodoi{10.1086/378200}

\bibitem[{{Sofue}(2009)}]{2009PASJ...61..153S}
{Sofue}, Y. 2009, \pasj, 61, 153, \dodoi{10.1093/pasj/61.2.153}

\bibitem[{{Sofue}(2013)}]{2013PASJ...65..118S}
---. 2013, \pasj, 65, 118, \dodoi{10.1093/pasj/65.6.118}

\bibitem[{{Sofue}(2017)}]{2017PASJ...69R...1S}
---. 2017, \pasj, 69, R1, \dodoi{10.1093/pasj/psw103}

\bibitem[{{Sofue}(2020)}]{2020Galax...8...37S}
---. 2020, Galaxies, 8, 37, \dodoi{10.3390/galaxies8020037}

\bibitem[{{Sofue} \& {Rubin}(2001)}]{2001ARA&A..39..137S}
{Sofue}, Y., \& {Rubin}, V. 2001, \araa, 39, 137, \dodoi{10.1146/annurev.astro.39.1.137}

\bibitem[{{Springel} {et~al.}(2005){Springel}, {White}, {Jenkins}, {Frenk}, {Yoshida}, {Gao}, {Navarro}, {Thacker}, {Croton}, {Helly}, {Peacock}, {Cole}, {Thomas}, {Couchman}, {Evrard}, {Colberg}, \& {Pearce}}]{2005Natur.435..629S}
{Springel}, V., {White}, S. D.~M., {Jenkins}, A., {et~al.} 2005, \nat, 435, 629, \dodoi{10.1038/nature03597}

\bibitem[{{Springel} {et~al.}(2008){Springel}, {White}, {Frenk}, {Navarro}, {Jenkins}, {Vogelsberger}, {Wang}, {Ludlow}, \& {Helmi}}]{2008Natur.456...73S}
{Springel}, V., {White}, S.~D.~M., {Frenk}, C.~S., {et~al.} 2008, \nat, 456, 73, \dodoi{10.1038/nature07411}

\bibitem[{{Steigman} \& {Turner}(1985)}]{1985NuPhB.253..375S}
{Steigman}, G., \& {Turner}, M.~S. 1985, Nuclear Physics B, 253, 375, \dodoi{10.1016/0550-3213(85)90537-1}

\bibitem[{{Su} {et~al.}(2022){Su}, {Lin}, {Pan}, {L{\'o}pez Cob{\'a}}, {Hsieh}, {S{\'a}nchez}, {Thorp}, {Bureau}, \& {Ellison}}]{2022ApJ...934..173S}
{Su}, Y.-C., {Lin}, L., {Pan}, H.-A., {et~al.} 2022, \apj, 934, 173, \dodoi{10.3847/1538-4357/ac77fd}

\bibitem[{{The LZ Collaboration} {et~al.}(2015){The LZ Collaboration}, {Akerib}, {Akerlof}, {Akimov}, {Alsum}, {Ara{\'u}jo}, {Bai}, {Bailey}, {Balajthy}, {Balashov}, {Barry}, {Bauer}, {Beltrame}, {Bernard}, {Bernstein}, {Biesiadzinski}, {Boast}, {Bolozdynya}, {Boulton}, {Bramante}, {Buckley}, {Bugaev}, {Bunker}, {Burdin}, {Busenitz}, {Carels}, {Carlsmith}, {Carlson}, {Carmona-Benitez}, {Cascella}, {Chan}, {Cherwinka}, {Chiller}, {Chiller}, {Craddock}, {Currie}, {Cutter}, {da Cunha}, {Dahl}, {Dasu}, {Davison}, {de Viveiros}, {Dobi}, {Dobson}, {Druszkiewicz}, {Edberg}, {Edwards}, {Edwards}, {Elnimr}, {Emmet}, {Faham}, {Fiorucci}, {Ford}, {Francis}, {Fu}, {Gaitskell}, {Gantos}, {Gehman}, {Gerhard}, {Ghag}, {Gilchriese}, {Gomber}, {Hall}, {Harris}, {Haselschwardt}, {Hertel}, {Hoff}, {Holbrook}, {Holtom}, {Huang}, {Hurteau}, {Ignarra}, {Jacobsen}, {Ji}, {Ji}, {Johnson}, {Ju}, {Kamdin}, {Kazkaz}, {Khaitan}, {Khazov}, {Khromov}, {Konovalov}, {Korolkova}, {Kraus}, {Krebs}, {Kudryavtsev}, {Kumpan}, {Kyre}, {Larsen},
  {Lee}, {Lenardo}, {Lesko}, {Liao}, {Lin}, {Lindote}, {Lippincott}, {Liu}, {Liu}, {Lopes}, {Lorenzon}, {Luitz}, {Majewski}, {Malling}, {Manalaysay}, {Manenti}, {Mannino}, {Markley}, {Martin}, {Marzioni}, {McKinsey}, {Mei}, {Meng}, {Miller}, {Mock}, {Monzani}, {Morad}, {Murphy}, {Nelson}, {Neves}, {Nikkel}, {O'Neill}, {O'Dell}, {O'Sullivan}, {Olevitch}, {Oliver-Mallory}, {Palladino}, {Pangilinan}, {Patton}, {Pease}, {Piepke}, {Powell}, {Preece}, {Pushkin}, {Ratcliff}, {Reichenbacher}, {Reichhart}, {Rhyne}, {Rodrigues}, {Rose}, {Rosero}, {Saba}, {Sarychev}, {Schnee}, {Schubnell}, {Scovell}, {Shaw}, {Shutt}, {Silva}, {Skarpaas}, {Skulski}, {Solovov}, {Sorensen}, {Sosnovtsev}, {Stancu}, {Stark}, {Stephenson}, {Stiegler}, {Sumner}, {Sundarnath}, {Szydagis}, {Taylor}, {Taylor}, {Tennyson}, {Terman}, {Thomas}, {Thomson}, {Tiedt}, {To}, {Tom{\'a}s}, {Tripathi}, {Tull}, {Tvrznikova}, {Uvarov}, {Va'vra}, {van der Grinten}, {Verbus}, {Vuosalo}, {Waldron}, {Wang}, {Webb}, {Wei}, {While}, {White}, {Whitis}, {Wisniewski},
  {Witherell}, {Wolfs}, {Woods}, {Woodward}, {Worm}, {Yeh}, {Yin}, {Young}, \& {Zhang}}]{2015arXiv150902910T}
{The LZ Collaboration}, {Akerib}, D.~S., {Akerlof}, C.~W., {et~al.} 2015, arXiv e-prints, arXiv:1509.02910, \dodoi{10.48550/arXiv.1509.02910}

\bibitem[{{Vargya} {et~al.}(2022){Vargya}, {Sanderson}, {Sameie}, {Boylan-Kolchin}, {Hopkins}, {Wetzel}, \& {Graus}}]{2022MNRAS.516.2389V}
{Vargya}, D., {Sanderson}, R., {Sameie}, O., {et~al.} 2022, \mnras, 516, 2389, \dodoi{10.1093/mnras/stac2069}

\bibitem[{{Walter} {et~al.}(2008){Walter}, {Brinks}, {de Blok}, {Bigiel}, {Kennicutt}, {Thornley}, \& {Leroy}}]{2008AJ....136.2563W}
{Walter}, F., {Brinks}, E., {de Blok}, W.~J.~G., {et~al.} 2008, \aj, 136, 2563, \dodoi{10.1088/0004-6256/136/6/2563}

\bibitem[{{Watkins} {et~al.}(2022){Watkins}, {Salo}, {Laurikainen}, {D{\'\i}az-Garc{\'\i}a}, {Comer{\'o}n}, {Janz}, {Su}, {Buta}, {Athanassoula}, {Bosma}, {Ho}, {Holwerda}, {Kim}, {Knapen}, {Laine}, {Men{\'e}ndez-Delmestre}, {Peletier}, {Sheth}, \& {Zaritsky}}]{2022A&A...660A..69W}
{Watkins}, A.~E., {Salo}, H., {Laurikainen}, E., {et~al.} 2022, \aap, 660, A69, \dodoi{10.1051/0004-6361/202142627}

\bibitem[{{Weber} \& {de Boer}(2010)}]{2010A&A...509A..25W}
{Weber}, M., \& {de Boer}, W. 2010, \aap, 509, A25, \dodoi{10.1051/0004-6361/200913381}

\bibitem[{{Yasin} {et~al.}(2023){Yasin}, {Desmond}, {Devriendt}, \& {Slyz}}]{2023MNRAS.tmp.1363Y}
{Yasin}, T., {Desmond}, H., {Devriendt}, J., \& {Slyz}, A. 2023, \mnras, \dodoi{10.1093/mnras/stad1183}

\bibitem[{{York} {et~al.}(2000){York}, {Adelman}, {Anderson}, {Anderson}, {Annis}, {Bahcall}, {Bakken}, {Barkhouser}, {Bastian}, {Berman}, {Boroski}, {Bracker}, {Briegel}, {Briggs}, {Brinkmann}, {Brunner}, {Burles}, {Carey}, {Carr}, {Castander}, {Chen}, {Colestock}, {Connolly}, {Crocker}, {Csabai}, {Czarapata}, {Davis}, {Doi}, {Dombeck}, {Eisenstein}, {Ellman}, {Elms}, {Evans}, {Fan}, {Federwitz}, {Fiscelli}, {Friedman}, {Frieman}, {Fukugita}, {Gillespie}, {Gunn}, {Gurbani}, {de Haas}, {Haldeman}, {Harris}, {Hayes}, {Heckman}, {Hennessy}, {Hindsley}, {Holm}, {Holmgren}, {Huang}, {Hull}, {Husby}, {Ichikawa}, {Ichikawa}, {Ivezi{\'c}}, {Kent}, {Kim}, {Kinney}, {Klaene}, {Kleinman}, {Kleinman}, {Knapp}, {Korienek}, {Kron}, {Kunszt}, {Lamb}, {Lee}, {Leger}, {Limmongkol}, {Lindenmeyer}, {Long}, {Loomis}, {Loveday}, {Lucinio}, {Lupton}, {MacKinnon}, {Mannery}, {Mantsch}, {Margon}, {McGehee}, {McKay}, {Meiksin}, {Merelli}, {Monet}, {Munn}, {Narayanan}, {Nash}, {Neilsen}, {Neswold}, {Newberg}, {Nichol}, {Nicinski},
  {Nonino}, {Okada}, {Okamura}, {Ostriker}, {Owen}, {Pauls}, {Peoples}, {Peterson}, {Petravick}, {Pier}, {Pope}, {Pordes}, {Prosapio}, {Rechenmacher}, {Quinn}, {Richards}, {Richmond}, {Rivetta}, {Rockosi}, {Ruthmansdorfer}, {Sandford}, {Schlegel}, {Schneider}, {Sekiguchi}, {Sergey}, {Shimasaku}, {Siegmund}, {Smee}, {Smith}, {Snedden}, {Stone}, {Stoughton}, {Strauss}, {Stubbs}, {SubbaRao}, {Szalay}, {Szapudi}, {Szokoly}, {Thakar}, {Tremonti}, {Tucker}, {Uomoto}, {Vanden Berk}, {Vogeley}, {Waddell}, {Wang}, {Watanabe}, {Weinberg}, {Yanny}, {Yasuda}, \& {SDSS Collaboration}}]{2000AJ....120.1579Y}
{York}, D.~G., {Adelman}, J., {Anderson}, John~E., J., {et~al.} 2000, \aj, 120, 1579, \dodoi{10.1086/301513}

\bibitem[{{Zhang} {et~al.}(2013){Zhang}, {Rix}, {van de Ven}, {Bovy}, {Liu}, \& {Zhao}}]{2013ApJ...772..108Z}
{Zhang}, L., {Rix}, H.-W., {van de Ven}, G., {et~al.} 2013, \apj, 772, 108, \dodoi{10.1088/0004-637X/772/2/108}

\end{thebibliography}
\bibliographystyle{aasjournal}

\end{document}